\documentclass[reprint, onecolumn, aps, prd%
 amsmath,amssymb,
 aps, physrev,
]{revtex4-2}

\usepackage{graphicx}
\usepackage{dcolumn}
\usepackage{bm}

\usepackage[utf8]{inputenc}
\usepackage{amsmath,amssymb}
\usepackage{graphicx}
\usepackage{hyperref}
\usepackage{geometry}
\usepackage{centernot}
\usepackage{color}
\usepackage{natbib}
\usepackage{comment}

\begin{document}


\title{\textbf{Chemical and Kinetic Equilibrium in Cosmology: Facts and Myths} 
}%

\author{Stefano Profumo}
\affiliation{Department of Physics, University of California, Santa Cruz, CA 95064, USA}
\affiliation{Santa Cruz Institute for Particle Physics (SCIPP), Santa Cruz, CA 95064, USA}

\date{\today}

\begin{abstract}
\noindent We clarify that chemical and kinetic equilibration in the early Universe are distinct: neither implies the other, and the ordering of their decouplings need not be universal. 
We  illustrate this with Standard–Model neutrino decoupling, 
strong–washout leptogenesis, 
dark–matter scenarios where kinetic decoupling precedes chemical freeze–out (resonant/forbidden, con\-ver\-sion/co\-an\-ni\-hi\-lation, coscattering), and dark sectors at with temperatures distinct from the visible-sector temperature,  with semi–annihilation or $3\!\to\!2$ cannibal dynamics. 
The moral of the story is simple: Chemical equilibrium governs numbers, kinetic equilibrium governs shapes. In an expanding Universe the operators that control them rarely fade at the same time, and when they do not, the order of decoupling is model dependent. Turning to phase–space evolution whenever momentum selectivity matters is the surest way to obtain robust cosmological predictions.
\end{abstract}

\maketitle

\section{Introduction}

In the evolving universe, the interplay between chemical and kinetic equilibrium governs the abundance and distribution of matter across cosmic time. Understanding these equilibrium concepts is fundamental to cosmology, as they determine everything from primordial nucleosynthesis in the early universe to star formation in molecular clouds and the ionization structure of the intergalactic medium.

Chemical equilibrium is achieved when the rates of formation and destruction of a given species balance, resulting in a time-independent number density. For a reaction network involving species $i$, this condition can be expressed as $\frac{dn_i}{dt} = \sum_{\text{formation}} R_{f,i} - \sum_{\text{destruction}} R_{d,i} = 0$, where $R_{f,i}$ and $R_{d,i}$ represent the formation and destruction rates respectively \citep{Peebles1993, Weinberg2008}. The establishment of chemical equilibrium depends critically on the reaction timescales compared to the relevant dynamical timescales of the system. In cosmological contexts, this often involves comparing reaction rates to the Hubble expansion rate $H(t)$. When reaction timescales are much shorter than the expansion timescale, $\tau_{\text{reaction}} \ll H^{-1}$, chemical equilibrium is maintained \citep{Kolb1990}. For a generic reaction $A + B \leftrightarrow C + D$, chemical equilibrium implies the law of mass action,
\begin{equation}
\frac{n_C n_D}{n_A n_B} = K_{\text{eq}}(T),
\end{equation}
where $n_i$ are the number densities of the species and $K_{\text{eq}}(T)$ is the temperature-dependent equilibrium constant. The latter is determined by the ratio of the partition functions of the reactants and products and, in the dilute limit, depends exponentially on the relevant binding energies \citep{Shu1991}.

Kinetic equilibrium (also termed thermal equilibrium) refers to the establishment of a thermal velocity distribution for particles of a given species. This occurs when collision processes are sufficiently frequent to maintain the characteristic Maxwell-Boltzmann, Fermi-Dirac, or Bose-Einstein distributions appropriate for the particle statistics \citep{Bernstein1988, Dodelson2003}. The condition for kinetic equilibrium requires that the collision timescale be much shorter than other relevant timescales in the system: $\tau_{\text{collision}} = \frac{1}{n \langle \sigma v \rangle} \ll \tau_{\text{system}}$, where $n$ is the particle density, $\langle \sigma v \rangle$ is the thermally averaged collision cross section, and $\tau_{\text{system}}$ represents the characteristic evolution timescale (e.g., $H^{-1}$ in cosmology) \citep{Weinberg2008}. In kinetic equilibrium, particles maintain a thermal distribution characterized by a single temperature $T$, with all translational degrees of freedom sharing this temperature. This is a stronger, but distinct, condition than chemical equilibrium and typically requires higher densities and interaction rates.

When kinetic equilibrium is established, the number density of a species can be calculated using statistical mechanics principles. The general expression depends on the quantum statistics obeyed by the particles. For non-relativistic particles in the classical limit, where the thermal de Broglie wavelength $\lambda_T = h/\sqrt{2\pi m k T}$ satisfies $n \lambda_T^3 \ll 1$, the equilibrium number density is:
\begin{equation}
n = g \left(\frac{m k T}{2\pi \hbar^2}\right)^{3/2} \exp\left(\frac{\mu - m c^2}{k T}\right),
\label{eq:classical}
\end{equation}
where $g$ is the statistical degeneracy factor, { which counts the number of internal degrees of freedom of the species (such as spin or polarization states). For clarity, we stress here that $g$ always denotes the internal (statistical) degeneracy factor throughout this work.}, $m$ is the particle rest mass, $\mu$ is the chemical potential, and other symbols have their usual meanings \citep{Pathria2011}.

For fermions, the number density follows Fermi-Dirac statistics:
\begin{equation}
n = g \int \frac{d^3p}{(2\pi\hbar)^3} \frac{1}{\exp\left[\frac{E(p) - \mu}{kT}\right] + 1},
\label{eq:fermi}
\end{equation}
while for bosons the distribution follows Bose-Einstein statistics:
\begin{equation}
n = g \int \frac{d^3p}{(2\pi\hbar)^3} \frac{1}{\exp\left[\frac{E(p) - \mu}{kT}\right] - 1},
\label{eq:bose}
\end{equation}
where $E(p) = \sqrt{p^2c^2 + m^2c^4}$ is the relativistic energy-momentum relation \citep{Kolb1990, Dodelson2003}.

Several important asymptotic limits are relevant in cosmological applications. In the ultra-relativistic limit ($kT \gg mc^2$), when $\mu \ll kT$, both fermions and bosons exhibit simple power-law scaling with temperature:
\begin{equation}
n = \begin{cases}
\frac{3\zeta(3)}{4\pi^2} g \left(\frac{kT}{\hbar c}\right)^3 & \text{(fermions)} \\
\frac{\zeta(3)}{\pi^2} g \left(\frac{kT}{\hbar c}\right)^3 & \text{(bosons)}
\end{cases}
\end{equation}
where $\zeta(3) \approx 1.202$ is the Riemann zeta function \citep{Weinberg2008}. In the non-relativistic limit ($kT \ll mc^2$), the quantum distributions reduce to the classical form given by Eq.~\eqref{eq:classical} when quantum effects are negligible. Finally, when $\mu \gg kT$, fermions become degenerate with number density $n = \frac{g}{6\pi^2} \left(\frac{\mu}{\hbar c}\right)^3$ for $\mu \gg mc^2$.

These equilibrium expressions form the foundation for calculating particle abundances in various cosmological epochs, from the radiation-dominated early universe through structure formation to the present day \citep{Peebles1993, Weinberg2008}. However, the relationship between chemical and kinetic equilibrium in cosmological contexts is more nuanced than often assumed in the literature. This paper addresses a few key questions that challenge common misconceptions about the connection between these equilibrium concepts, broadly around the question of whether chemical decoupling always precede kinetic decoupling:

\begin{enumerate}
\item Does early chemical equilibrium imply a thermal velocity distribution at later times?
\item Must a species that \textit{is} in chemical equilibrium be in kinetic equilibrium with the Standard Model plasma?
\item If the former does not hold, would chemical equilibrium imply kinetic equilibrium  within a dark sector with its own temperature?
\end{enumerate}

We demonstrate that the answer to all the above questions is no, through detailed analysis of concrete examples from neutrino physics, dark matter scenarios, and beyond-Standard Model cosmology. These scenarios illuminate the rich phenomenology possible when chemical and kinetic processes operate on different timescales, commonly encountered in beyond-Standard Model cosmology and dark sector physics. We provide a general framework for determining when chemical and kinetic equilibrium end, and  develop detailed examples and models that exemplify these phenomena, analyze their common underlying physics, and explore some of their quantitative observational consequences.
\section{General Formalism and Notation}
This section collects the kinematic conventions, kinetic equations, rate definitions, and analysis tools used throughout the paper. {We aim to make this section fully self-contained by clearly defining every symbol and introducing standard references.}

\subsection{Phase space, Liouville flow, and comoving variables}
We describe each species $i$ by its phase–space density $f_i(p, t)$ obeying, in a homogeneous FRW background with scale factor $a(t)$,
\begin{equation}\label{eq:boltzmann-master}
(\partial_t -H \, p\cdot\nabla_p) f_i(p, t) = C_i[f] , \qquad H \equiv \dot a/a ,
\end{equation}
where $H$ is the Hubble expansion rate. {Here $C_i[f]$ is the collision operator describing interactions. We use natural units $\hbar = c = k_B = 1$ throughout. For general background on Boltzmann equations in cosmology, see Refs.~\cite{Kolb1990, Gondolo2012}.}

It is often convenient to work with comoving momentum $q \equiv a p$ or with the dimensionless variable $y \equiv p/T_{\rm cm}$, where $T_{\rm cm} \propto a^{-1}$ is any reference comoving temperature (e.g. the photon temperature in the absence of entropy dumps). The number, energy, and pressure moments are
\begin{equation}
n_i = \frac{g_i}{(2\pi)^3}\int d^3p f_i(p), \qquad
\rho_i = \frac{g_i}{(2\pi)^3}\int d^3p E_i(p) f_i(p), \qquad
P_i = \frac{g_i}{(2\pi)^3}\int d^3p \frac{p^2}{3E_i(p)} f_i(p),
\end{equation}
with $E_i = \sqrt{p^2 + m_i^2}$ and internal degeneracy $g_i$. {The factor $g_i$ counts the number of internal spin, polarization, or color degrees of freedom.}

\subsection{Collision operator split and equilibrium concepts}
We separate the collision operator $C_i$ into inelastic (number–changing) and elastic (momentum–exchange) pieces,
\begin{equation}
C_i[f] = C^{\rm inel}_i[f] + C^{\rm el}_i[f],
\end{equation}
where $C^{\rm inel}$ contains decays/inverse decays and $2\leftrightarrow2$, $3\leftrightarrow2$, ... reactions that change $\sum_i N_i$, while $C^{\rm el}$ contains scatterings that conserve particle number but exchange momentum and energy. {This decomposition is standard in kinetic theory; see Ref.~\cite{Dolgov:2002wy} for neutrino transport and Ref.~\cite{BringmannHofmann2007} for dark matter applications.}

\paragraph{Chemical equilibrium.} For a given reaction network $\sum_j \, \nu_{aj} X_j \leftrightarrow 0$, chemical equilibrium imposes constraints on chemical potentials, $\sum_j \nu_{aj}\mu_j = 0$ for all active processes $a$. Operationally, letting $\Gamma_{\rm chem}$ denote the slowest relevant per-particle number–changing rate, chemical equilibrium obtains when
\begin{equation}
\Gamma_{\rm chem}(T) \gg H(T).
\end{equation}
{In practice, $\Gamma_{\rm chem}$ can represent annihilation, decay, or conversion rates. As long as these processes are much faster than Hubble expansion, number densities follow their equilibrium values.}

\paragraph{Kinetic equilibrium.} Kinetic equilibrium demands that momentum exchange be sufficiently fast to maintain a thermal distribution for $f_i$, typically Maxwell–Boltzmann/Fermi–Dirac/Bose–Einstein with some $(T_i, \mu_i)$. A robust diagnostic is the momentum–transfer rate $\gamma_p$, defined below. Kinetic equilibrium requires
\begin{equation}
\gamma_p^{(X)}(p) \gg H \qquad \text{for the momenta that dominate $n_i$ and $\rho_i$},
\end{equation}
where the superscript $(X)$ indicates the bath mediating kineticization (e.g. Standard Model (SM), or Dark Sector (DS)). {Because of transport weighting (see the discussion in sec.~\ref{sec:transport} below), high-$p$ modes typically decouple first. This diagnostic has been extensively used in studies of dark matter kinetic decoupling~\cite{Boehm:2000gq, BringmannHofmann2007}.} {For distribution-level studies of elastic scattering in both freeze-in and freeze-out scenarios, see Ref.~\cite{Du:2021jcj}. In agreement with our argument, it was shown that a sufficiently large elastic scattering rate can establish kinetic equilibrium even during freeze-in, where the condition for chemical equilibrium is never satisfied.}

\subsection{Reaction rates and thermally averaged kernels}
For $2\leftrightarrow2$ inelastic reactions $a+b\leftrightarrow c+d$,
\begin{align}
\Gamma_{a,\,{\rm inel}} &= \sum_b n_b \,\langle \sigma v_{\rm M\text{\o} l}\rangle_{ab\to \cdots}\,,
&
\langle \sigma v_{\rm M\text{\o} l}\rangle &= \frac{\int d\Pi_a d\Pi_b\, f_a f_b\, \sigma\, v_{\rm M\text{\o} l}}{\int d\Pi_a d\Pi_b\, f_a f_b}\,,
\end{align}
with Lorentz-invariant measure $d\Pi = g \, d^3p /[(2\pi)^3 2E]$ and Møller velocity $v_{\rm M\o l}$. {The thermal average $\langle \sigma v_{\rm M\o l}\rangle$ ensures relativistic invariance and is standard in relic density calculations~\cite{Gondolo2012}.}

For decays/inverse decays $X \leftrightarrow ij$,
\begin{equation}
\Gamma_{X,\,\rm chem} \simeq \Gamma_{X\to ij} \times \frac{n_X}{n_X^{\rm (eq)}(T)} \, ,
\end{equation}
with finite-density effects included when needed.

For $3\to2$ reactions in a self-thermalized sector,
\begin{equation}
\Gamma_{3\to2} \equiv n^2 \langle \sigma v^2 \rangle_{3\to2}.
\end{equation}

During radiation domination,
\begin{equation}
H(T) \simeq 1.66 \, g_\star^{1/2}(T) \frac{T^2}{M_{\rm Pl}},
\end{equation}
where {$g_\star(T)$ is the effective number of relativistic degrees of freedom contributing to the energy density~\cite{Kolb1990}, and $M_{\rm Pl} \simeq 1.22\times10^{19}$ GeV is the (non-reduced) Planck mass.}

\subsection{Transport (momentum–exchange) rate and heat exchange}\label{sec:transport}
Elastic $2\!\to\!2$ scattering $i\,t\to i\,t$ contributes to \emph{momentum} and \emph{heat} exchange. The appropriate cross section is transport-weighted,
\begin{equation}
\sigma_{\rm mt}(s) \equiv \int d\Omega\,(1-\cos\theta)\,\frac{d\sigma}{d\Omega}\,,
\end{equation}
which suppresses forward scattering. {This “transport weighting” ensures that scatterings which only deflect the incoming particle by a very small angle, while perhaps dominating the total elastic cross section, contribute negligibly to momentum transfer. In other words, the factor $(1-\cos\theta)$ penalizes forward scattering, so that $\sigma_{\rm mt}$ correctly captures the rate at which the momentum of particle $i$ is randomized through interactions with the target bath $t$ (see e.g. Refs.~\cite{Hofmann:2001bi,Boehm:2000gq,Vogelsberger:2015gpr}).}

For a heavy species $i$ scattering on relativistic targets $t$ with temperature $T_t$,
\begin{equation}
\gamma_p^{(t)}(i)\ \simeq\ n_t\,\big\langle \sigma_{\rm mt}\, v \big\rangle\,\frac{T_t}{m_i}\,,
\label{eq:gamma-p}
\end{equation}
up to ${\cal O}(1)$ factors. {Here $n_t$ is the number density of target particles, $\langle \sigma_{\rm mt} v \rangle$ is the thermal average of the transport cross section, $T_t$ is the bath temperature, and $m_i$ is the mass of the heavy scattering particle. The factor $T_t/m_i$ reflects the fractional momentum transfer per collision in the heavy–light limit. This rate $\gamma_p$ is often called the ``momentum transfer rate'' or ``drag rate'' in the context of dark matter–baryon or dark matter–radiation interactions (see, e.g., Refs.~\cite{BringmannHofmann2007,Slatyer:2018aqg}).}

A corresponding \emph{heat-exchange} (energy-transfer) rate $\gamma_E$ can be defined analogously, and in small-angle regimes $\gamma_E \sim \gamma_p$ (precise proportionality depends on kinematics). {In practice, the ratio $\gamma_E/\gamma_p$ depends on how efficiently kinetic energy is exchanged relative to momentum exchange; for example, in Rutherford-like scattering with light targets the two are nearly proportional, whereas for massive targets the proportionality can differ. Both $\gamma_p$ and $\gamma_E$ must be computed including plasma effects such as Debye screening, finite-temperature masses, and Coulomb logarithms (see e.g. Refs.~\cite{BringmannHofmann2007,Slatyer:2018aqg}). These refinements are essential whenever long-range mediators or gauge interactions are involved.}

{In summary, $\gamma_p$ quantifies the rate at which the bulk momentum distribution of a species is isotropized, while $\gamma_E$ controls the rate of thermal energy exchange with the bath. Kinetic equilibrium is maintained as long as $\gamma_p/H \gg 1$ at the momenta that dominate the number and energy densities. Once $\gamma_p/H$ falls below unity, elastic scattering is no longer efficient enough to sustain a Maxwell–Boltzmann (or Fermi–Dirac/Bose–Einstein) distribution, and the system kinetically decouples.}

{For numerical treatments that go beyond such analytical approaches, see e.g.\ Ref.~\cite{Binder:2021bmg}, where the public code {\sc DRAKE} was introduced as a dedicated tool to follow departures from local thermal equilibrium at the distribution level.}

\subsection{Free-streaming solution after kinetic decoupling}
If at some time $t_{\rm kd}$ the appropriate $\gamma_p/H$ falls below unity across populated momenta and inelastic terms are negligible thereafter, Eq.~(6) integrates along characteristics to
\begin{equation}
f_i(p, t) = f_i\left(\frac{a(t)}{a(t_{\rm kd})} p, t_{\rm kd}\right) + \int_{t_{\rm kd}}^t dt' S^{\rm inel}_i\left(\frac{a(t)}{a(t')} p, t'\right),
\label{eq:free-stream}
\end{equation}
where $S^{\rm inel}$ collects any subsequent injection/depletion sources. {Thus the post-decoupling spectrum is a redshifted memory of the distribution at kinetic decoupling, plus any later injection; no single $(T,\mu)$ need describe it. This statement underlies free-streaming effects of neutrinos~\cite{Dolgov:2002wy} and warm dark matter~\cite{Bode:2000gq}.}

\subsection{Two-bath cosmologies and temperatures}
When a dark sector communicates feebly with the SM, it is useful to track an SM temperature $T$ and a DS temperature $T_d$ with $T_d/T$ evolving by entropy exchange. Portal energy flow $Q_{\rm portal}$ enters the temperature evolution via
\begin{align}
\dot\rho_{\rm SM} + 3H(\rho_{\rm SM} + P_{\rm SM}) &= +Q_{\rm portal},\\
\dot\rho_{\rm DS} + 3H(\rho_{\rm DS} + P_{\rm DS}) &= -Q_{\rm portal},
\end{align}
with $Q_{\rm portal}$ obtained from the appropriate collision integrals. Within the DS, species may share $T_d$ (fast intra-DS elastic) or carry distinct kinetic temperatures $T_i$ (marginal intra-DS elastic), in which case one evolves separate heat equations. 
{Such two-temperature setups are standard in hidden-sector cosmology~\cite{Chu:2011be, Feng2008}.} {Explicit examples of two-component dark sectors and the impact of DM self-scatterings on equilibration were analyzed numerically in Ref.~\cite{Hryczuk:2022gay}.}

\subsection{Exact collision integrals vs. Fokker–Planck/Langevin}
\noindent \paragraph{Full collision integrals (``fBE'').} Exact $2\!\leftrightarrow\!2$ kernels with quantum statistics and (when needed) $3\!\leftrightarrow\!2$ are mandatory when (i) annihilation/production is sharply momentum-selective (narrow resonances, forbidden/threshold kinematics), (ii) large-angle scatterings matter, or (iii) precision neutrino transport is required. Efficient formulations reduce the nine-dimensional integrals to two-dimensional kernels in $y=p/T_{\rm cm}$.\\

\noindent \paragraph{Fokker–Planck (FP) / Langevin.} In small-angle regimes (forward $t$-channel exchange), elastic scattering is accurately captured by an FP operator,
\begin{equation}
C^{\rm FP}_{\rm el}[f] \;=\; \partial_{p_i}\!\left[A_i(p)\,f + \partial_{p_j}\!\left(B_{ij}(p)\,f\right)\right],
\label{eq:FP}
\end{equation}
with drag $A_i$ and diffusion $B_{ij}$ matched to the exact momentum \emph{and} heat exchange (thereby reproducing $\gamma_p$ and $\gamma_E$). A Langevin limit is efficient for heavy species in a relativistic bath. FP/Langevin should be benchmarked against full kernels near thresholds or when large-angle scatterings contribute appreciably. {Full Boltzmann integrals are mandatory in cases of sharp momentum selectivity, while Fokker–Planck approximations apply in forward-scattering regimes. For applications to neutrino transport and DM kinetic decoupling, see Refs.~\cite{Dolgov:2002wy, BringmannHofmann2007}.}

\subsection{Moment hierarchies and controlled closures}
Integrating Eq.~\eqref{eq:boltzmann-master} yields hierarchies for $n$, $\rho$, $P$, \dots. A widely used controlled closure retains number density and a kinetic temperature $T_i$ (or second moment); schematically,
\begin{align}
\dot n_i + 3H n_i &= \left[\text{inelastic sources/sinks (decays, (co)ann., $3\!\to\!2$)}\right], \label{eq:cbe-n}\\
\frac{3}{2}\left(\dot n_i T_i + n_i \dot T_i\right) + 5H n_i T_i &= -\,\mathcal{Q}_{\rm el}\big(T_i - T_{\rm bath}\big) + \mathcal{Q}_{\rm inel}\,,
\label{eq:cbe-T}
\end{align}
where $\mathcal{Q}_{\rm el}\propto \gamma_p$ and $\mathcal{Q}_{\rm inel}$ accounts for energy nonconservation per particle in number–changing reactions (e.g.\ self-heating from $3\!\to\!2$, recoil in inverse decays). This closure is accurate if $f_i$ remains close to Maxwellian; when high-$p$ tails or narrow bands dominate chemistry, one must resort to targeted $f$-space evolution. {This controlled closure is accurate as long as distributions remain close to Maxwellian. Deviations require momentum-resolved treatments, as emphasized in~\cite{BringmannHofmann2007, Binder:2017rgn}.}

\subsection{Chemical networks and chemical potentials}
For a network of reactions $\sum_j \nu_{aj}X_j \leftrightarrow 0$ in chemical equilibrium, solve $\sum_j \nu_{aj}\mu_j=0$ for $\{\mu_j\}$ (with conserved charges as Lagrange multipliers). For example,
\begin{itemize}
\item semi-annihilation $\chi\chi\leftrightarrow \chi\phi$ implies $2\mu_\chi=\mu_\chi+\mu_\phi$;
\item $3\!\leftrightarrow\!2$ cannibalism $\chi\chi\chi \leftrightarrow \chi\chi$ implies $3\mu_\chi=2\mu_\chi$ so $\mu_\chi=0$ in the chemically equilibrated DS (if only $\chi$ carries number);
\item coannihilation/conversions enforce $\mu_\chi\simeq\mu_\psi$.
\end{itemize}
When $\Gamma_{\rm chem}\gg H$ but $\gamma_p\lesssim H$, these relations hold at the level of \emph{number} densities while $f_i$ need not be thermal. {Examples include semi-annihilation $\chi\chi\leftrightarrow \chi\phi$, 3$\leftrightarrow$2 cannibal processes, and coannihilation/conversions, see~\cite{DEramo:2010keq, PappadopuloRuderman2016}.} {For conversion-driven freeze-out, distribution-level studies have recently been carried out in Ref.~\cite{Chatterjee:2025vdz}, complementing earlier moment-level analyses of semi-annihilations and cannibal processes.}


\section{Example Setup: Decays/Inverse Decays Without Elastic Scattering}
\label{sec:example}

To make the distinction between chemical and kinetic equilibrium explicit, 
consider a toy model in which a species $X$, injected, in general, out of thermal equilibrium, interacts with a thermal bath only via
number-changing decays and inverse decays
\begin{equation}
X \;\longleftrightarrow\; Y + Z,
\end{equation}
where $Y$ and $Z$ are lighter particles in kinetic and chemical equilibrium with the bath.
We assume that \emph{elastic} scattering processes involving $X$,
\begin{equation}
X + \text{SM} \to X + \text{SM},
\end{equation}
are negligible.\\

\noindent \paragraph{Chemical equilibrium.}  
The decay and inverse-decay rates $\Gamma_{\rm dec}$ are taken to satisfy 
$\Gamma_{\rm dec} \gg H$, so that the integrated number density $n_X$ 
closely follows its equilibrium value $n_X^{\rm eq}(T)$.  
For every $X$ removed by decay, an inverse decay promptly produces another,
ensuring detailed balance in the \emph{number} of $X$ particles.\\

\noindent \paragraph{Loss of kinetic equilibrium.}  
The momentum distribution $f_X(p,t)$, however, is determined entirely by the 
kinematics of the inverse decays, which inject $X$ particles at a 
characteristic momentum $p_0$ in the bath rest frame.  
Without frequent elastic scatterings to redistribute momenta, the spectrum 
does not relax to the Maxwell--Boltzmann form even though $n_X = n_X^{\rm eq}$.  
This is the simplest realisation of the hierarchy 
\[
\Gamma_{\rm inelastic} \equiv \Gamma_{\rm dec} \gg H \gg \Gamma_{\rm elastic}.
\]

\noindent \paragraph{Including elastic scattering.}  
If we now add an elastic scattering term with rate $\gamma$, the steady-state distribution 
becomes a weighted average of the injection spectrum and the Maxwell--Boltzmann spectrum,
\begin{equation}
f_\infty(p) \;=\; \frac{S(p) + \gamma f_{\rm MB}(p)}{\Gamma_{\rm dec} + \gamma}.
\end{equation}
When $\gamma \to 0$, we recover the purely non-thermal steady state;  
when $\gamma \gg \Gamma_{\rm dec}$, we recover kinetic equilibrium.\\

\begin{figure}[t]
    \centering
    \includegraphics[width=0.7\textwidth]{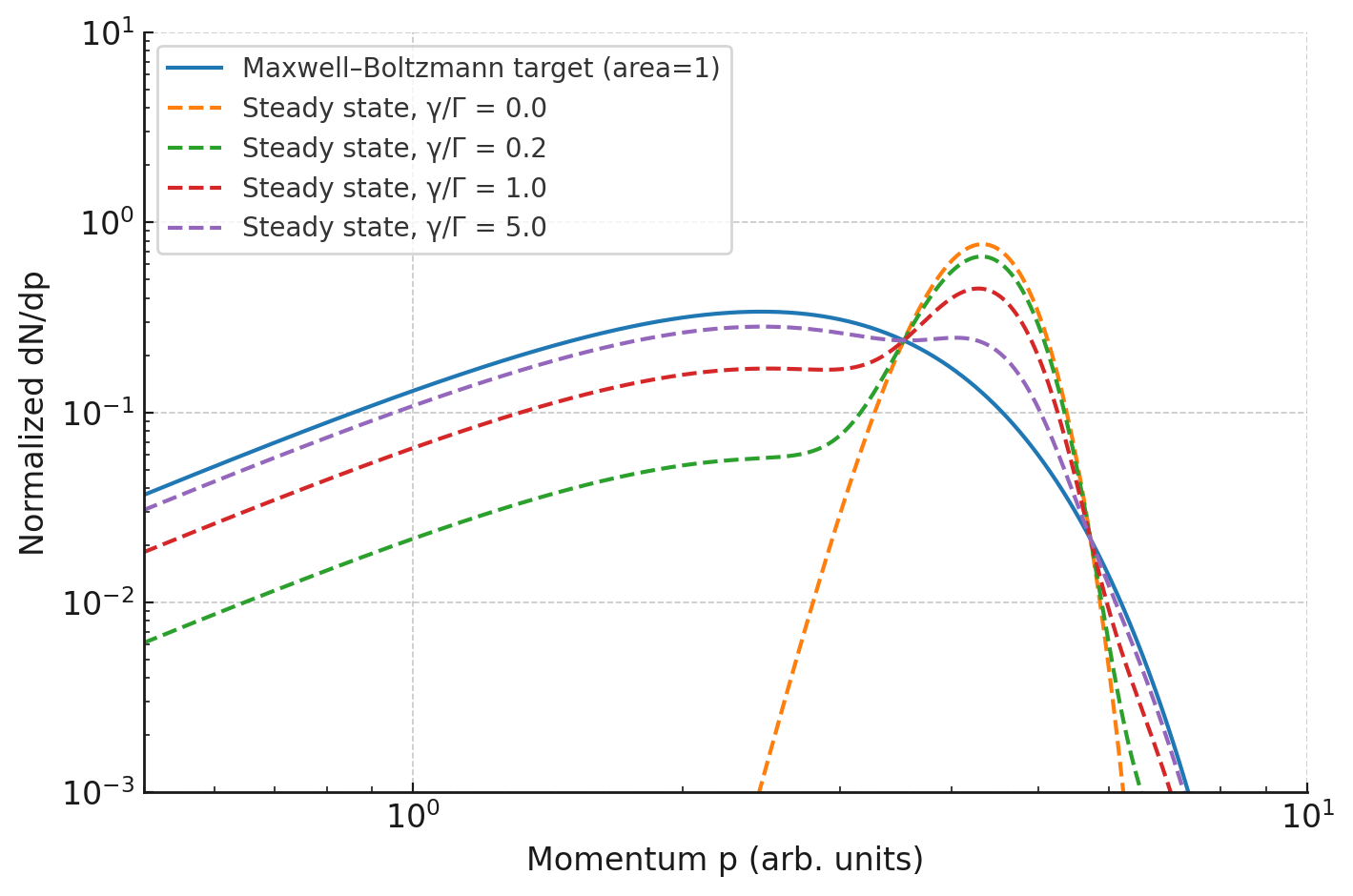}
    \caption{Steady-state number spectra for various elastic scattering rates $\gamma/\Gamma_{\rm dec}$.
    All spectra are normalized to the same total number (chemical equilibrium), 
    but only for large $\gamma$ does the shape match the Maxwell--Boltzmann distribution (kinetic equilibrium).}
    \label{fig:steadystate}
\end{figure}

\begin{figure}[t]
    \centering
    \includegraphics[width=0.7\textwidth]{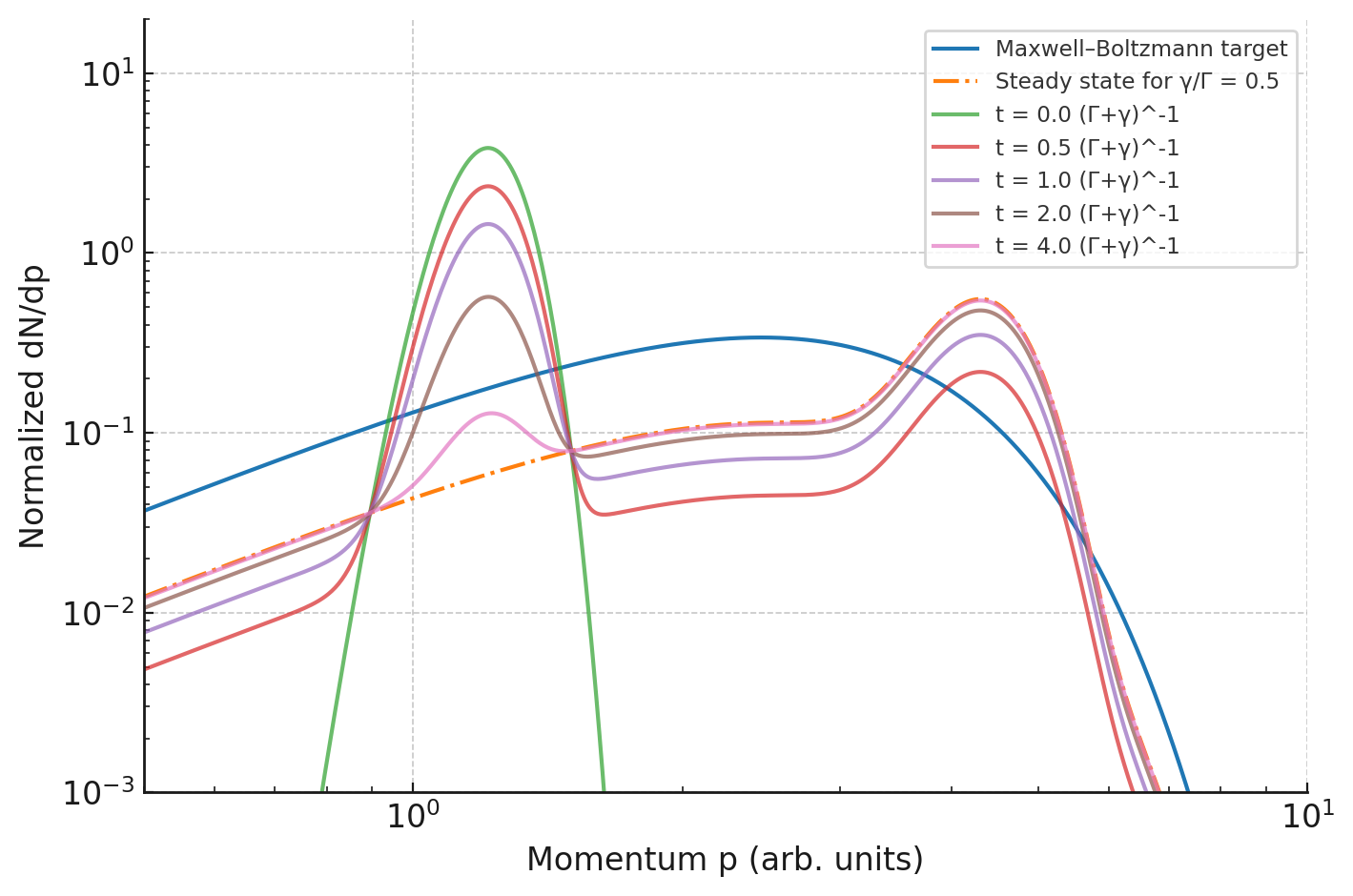}
    \caption{Time evolution toward the steady state for $\gamma/\Gamma_{\rm dec} = 0.5$.
    The distribution starts far from thermal (gree line) and converges to a mixed non-thermal/thermal shape
    over a timescale $(\Gamma_{\rm dec} + \gamma)^{-1}$, with the total number fixed at the
    equilibrium value throughout.}
    \label{fig:timedep}
\end{figure}

\noindent \paragraph{Numerical illustration.}  
Figures~\ref{fig:steadystate} and \ref{fig:timedep} show this explicitly for a 
choice of parameters where the initial non-thermal spectrum is a narrow Gaussian 
centered away from the thermal peak (shown in green in fig.~\ref{fig:timedep}).  
We plot the \emph{number spectrum} $dN/dp \propto p^2 f(p)$ on log--log scales, 
normalized so that the area matches that of the Maxwell--Boltzmann spectrum 
(chemical equilibrium).  
The first plot varies $\gamma$ to show the approach to kinetic equilibrium;  
the second shows the time evolution toward the steady state for $\gamma / \Gamma_{\rm dec} = 0.5$, in units of the timescale $(\Gamma_{\rm dec}+\gamma)^{-1}$.\\

The triple-peaked momentum spectrum in fig.~\ref{fig:timedep} results from the interplay between
(i) elastic scattering with the thermal bath, which drives the distribution
towards Maxwell--Boltzmann form, and (ii) inverse decays, which inject particles
with a characteristic momentum. The left-most, green peak is the initial $t=0$ non-thermal injection spectrum, assumed to be a Gaussian centered below the thermal peak, which vanishes in the steady-state regime shown in fig.~\ref{fig:steadystate}.\\

\noindent {\bf Thermal peak.}
Elastic scattering tends to establish a Maxwell--Boltzmann distribution,
\begin{equation}
\frac{dN_{\rm MB}}{dp} \;\propto\; p^{2}\, e^{-p^{2}/(2 m T)} ,
\end{equation}
which reaches its maximum at
\begin{equation}
p_{\rm MB} = \sqrt{2 m T}.
\end{equation}
For the parameters used in the figure ($m=3$, $T=1$), this gives
\begin{equation}
p_{\rm MB} \simeq \sqrt{6} \simeq 2.45,
\end{equation}
corresponding to the left-hand peak.\\

\noindent {\bf Injection peak.}
Inverse decays produce particles with a narrow momentum distribution,
\begin{equation}
S(p) \;\propto\; \exp\!\left[-\frac{(p-p_{0})^{2}}{2\sigma^{2}}\right],
\end{equation}
so that the number spectrum is
\begin{equation}
\frac{dN_{\rm inj}}{dp} \;\propto\; p^{2}\,S(p).
\end{equation}
Maximizing $\ln(p^{2} S(p))$ yields the location of the injection peak,
\begin{equation}
\frac{2}{p} - \frac{p - p_{0}}{\sigma^{2}} = 0
\quad\Rightarrow\quad
p_{\rm inj} = \frac{p_{0} + \sqrt{p_{0}^{2} + 8\sigma^{2}}}{2}.
\end{equation}
For the numerical choices $p_{0}=2.5\sqrt{mT} \simeq 4.33$ and
$\sigma = 0.3 \sqrt{mT} \simeq 0.52$, one finds
\begin{equation}
p_{\rm inj} \simeq 4.45,
\end{equation}
corresponding to the right-hand peak.\\

\noindent{\bf Combined spectrum.}
The full steady-state distribution is a weighted average of the thermal
and injection contributions,
\begin{equation}
\left.\frac{dN}{dp}\right|_{\infty}
= \frac{S(p) \;+\; \gamma \,\frac{dN_{\rm MB}}{dp}}{\Gamma+\gamma},
\end{equation}
where $\Gamma$ is the decay/inverse-decay rate and $\gamma$ the elastic
scattering rate.  When $\Gamma$ and $\gamma$ are comparable, both peaks
are clearly visible: the thermal bump at $p\simeq 2.45$ and the injection
bump at $p\simeq 4.45$.


\section{Does past chemical equilibrium enforce thermal velocities at later times?}
\label{sec:past-chem-not-thermal}

\noindent {\bf Claim:}
A species that \emph{was} in chemical equilibrium at an earlier epoch need not possess a thermal (Maxwell–Boltzmann/Fermi–Dirac/Bose–Einstein) velocity distribution at later times, whether or not chemical decoupling happens prior to kinetic decoupling.

\subsection*{Kinetic rationale}
Let $f(p,t)$ be the phase–space density of the species. Its evolution obeys Eq.~\eqref{eq:boltzmann-master}. Chemical equilibrium at time $t_\star$ constrains the \emph{zeroth moment} (abundance) via
$\Gamma_{\rm chem}(t_\star)\gg H(t_\star)$, but it does not determine the future \emph{shape} of $f$.
If, at some later time $t_{\rm kd}$, the momentum–transfer (transport) rate
$\gamma_p\!\sim\!n_{\rm bath}\langle\sigma_{\rm mt} v\rangle\,T_{\rm bath}/m$
falls below the Hubble rate for populated momenta, $\gamma_p(p)\ll H$, elastic scattering no longer maintains a thermal form. Neglecting $C_{\rm el}$ afterwards, \eqref{eq:boltzmann-master} integrates to Eq.~\eqref{eq:free-stream}  where as noted above $\mathcal{S}_{\rm inel}$ represents inelastic sources/sinks per comoving momentum.
Equation~\eqref{eq:free-stream} shows that the post–kinetic–decoupling spectrum is a redshifted memory of the shape at $t_{\rm kd}$, plus any later injection/depletion; neither operation forces a return to a single-temperature form. In particular, high-$p$ modes typically violate $\gamma_p(p)\gg H$ first (transport weighting $(1-\cos\theta)$ penalizes forward scatterings), so deviations from thermality seed from the top of the distribution and grow with time. We show two examples drawn from neutrino physics where chemical decoupling precedes/follows kinetic decoupling, and where the asymptotic momentum distribution is not thermal.\\[1cm]


\subsection{Standard-Model neutrinos (MeV epoch)}
\label{subsec:nu-MeV-expanded}

\noindent \noindent \paragraph{Setup and physical scales.}
Around the MeV epoch the active neutrinos $\nu=\{\nu_e,\nu_\mu,\nu_\tau\}$ depart from full equilibrium with the electromagnetic (EM) plasma $\{e^\pm,\gamma\}$. Hubble expansion during radiation domination is
$H\simeq 1.66\,g_*^{1/2}T^2/M_{\rm Pl}$,
while weak interaction rates scale roughly as $\sim G_F^2 T^5$ up to order-unity angular/energy weights. Finite-temperature QED corrections modify the EM equation of state and electron mass around $T\sim m_e$, and three-flavor oscillations redistribute energy and distortions among flavors. Momentum-resolved transport (Boltzmann or QKE) with exact $2\!\leftrightarrow2$ weak kernels, finite-$T$ QED thermodynamics, and oscillations provides the modern baseline \cite{Mangano2005,Escudero2019,AkitaYamaguchi2020,CieloEtAl2023}.\\

\noindent \noindent \paragraph{Chemical vs.\ kinetic rate hierarchy.}
Two operator classes in the weak collision term behave differently as $T$ falls through $m_e$:
\begin{itemize}
\item \emph{Chemical (number-changing):} $\nu\bar\nu \leftrightarrow e^+e^-$ controls $\mu_\nu$ and the possibility of tracking an equilibrium abundance. Its per-particle rate scales as
$\Gamma_{\rm chem}\sim n_{e^\pm}\langle\sigma v\rangle_{\nu\bar\nu\leftrightarrow e^+e^-}\propto n_{e^\pm} G_F^2 T^2$.
As $T\lesssim m_e$, $n_{e^\pm}$ drops exponentially and $\Gamma_{\rm chem}/H$ falls fastest.
\item \emph{Kinetic (momentum/energy exchange):} elastic $\nu$–$e^\pm$ scattering,
$\nu e^\pm \leftrightarrow \nu e^\pm$, continues to exchange momentum/heat with the EM bath with a transport rate
\begin{equation}
\gamma_p(p)\;\simeq\; n_{e^\pm}\,\big\langle \sigma_{\rm mt}(\nu e)\,v\big\rangle \,\frac{T}{E_\nu}\,,
\qquad
\sigma_{\rm mt}\equiv\!\int d\Omega\,(1-\cos\theta)\frac{d\sigma}{d\Omega},
\end{equation}
which, although it also weakens as $n_{e^\pm}$ declines, typically remains $\gtrsim H$ for a while \emph{after} pair processes have become inefficient. Neutrino–neutrino scatterings help redistribute within the $\nu$ sector but cannot by themselves maintain kinetic coupling to the EM bath.
\end{itemize}
The generic ordering during decoupling is therefore
\[
\Gamma_{\rm chem}\ \downarrow\ \text{first} \quad \text{and} \quad \gamma_p\ \downarrow\ \text{later},
\]
so that chemical decoupling precedes kinetic decoupling.\\

\noindent \noindent \paragraph{Momentum-space consequences.}
Solving
\begin{equation}
(\partial_t - H\,p\,\partial_p) f_\nu(p,t)
\;=\; C_{\rm ann}[f_\nu,f_{e^\pm}] \;+\; C_{\rm el}[f_\nu,f_{e^\pm}] \;+\; C_{\rm osc}[\,\varrho\,],
\end{equation}
where $C_{\rm osc}$ captures the collision operator associated with neutrino oscillations, reveals small, irreducible spectral distortions in $f_\nu$:
(i) high-$p$ modes (in comoving $y\equiv p/T$) remain more strongly coupled and receive a larger share of the late EM entropy flow, generating \emph{larger} deviations there;
(ii) flavor oscillations redistribute these distortions, largely equilibrating $\nu_e$ and $\nu_{\mu,\tau}$ features by the end of $e^\pm$ annihilation.
No single $(T_\nu,\mu_\nu)$ can exactly reproduce the momentum dependence of the relic spectra; a two-parameter ansatz misses the few$\times 10^{-3}$–$10^{-2}$ level $y$-dependent structure seen in full solutions \cite{Mangano2005,Escudero2019,AkitaYamaguchi2020,CieloEtAl2023}.\\

\noindent \noindent \paragraph{Entropy/energy flow and timing.}
Because $\nu$–$e^\pm$ scattering persists \emph{into} the $e^\pm$ annihilation epoch, part of the entropy released from $e^\pm\to\gamma$ is transferred to neutrinos. Momentum-resolved computations consistently find significant EM$\leftrightarrow\nu$ energy exchange well below the naive “instantaneous” decoupling temperature, extending to $T\sim 0.1$–$0.2$ MeV in the EM bath. This non-instantaneous reheating of neutrinos is exactly what produces the final-state distortions and the small upward shift in the radiation density.


\subsection{Heavy right-handed neutrinos with strong washout}
\label{subsec:N-strong-washout}

\noindent\paragraph{Setup and rate hierarchy.}
In thermal leptogenesis from a type-I seesaw, heavy Majorana neutrinos $N$ (mass $M_N$) couple to lepton doublets $L$ and the Higgs $H$ via
$\mathcal{L}\supset -y_N\,\overline{N}\,L H + \text{h.c.}$
The dominant \emph{inelastic} processes are decays and inverse decays,
\begin{equation}
N \leftrightarrow L H\,,
\qquad
\Gamma_D^0 \equiv \frac{|y_N|^2 M_N}{8\pi}\,,
\end{equation}
with finite-density/statistical corrections at $T\sim M_N$.
The strong-washout regime is characterized by
\begin{equation}
K \;\equiv\; \frac{\Gamma_D}{H}\Big|_{T=M_N} \;=\; \frac{|y_N|^2 M_N}{8\pi}\,\frac{M_{\rm Pl}}{1.66\,g_*^{1/2} M_N^2}
\;\simeq\; \frac{\tilde m}{m_*}\ \gg\ 1\,,
\label{eq:K}
\end{equation}
so that inverse decays keep $n_N$ close to $n_N^{\rm(eq)}(T)$ around $T\sim M_N$ \cite{Giudice:2003jh} (``chemical equilibrium'').

By contrast, the \emph{kinetic} coupling of $N$ to the SM bath is governed by elastic $2\!\to\!2$ processes such as
$N\,\ell\!\to\!N\,\ell$ and $N\,H\!\to\!N\,H$ via $t$-channel exchange of $H$ or $\ell$.
These amplitudes scale as $y_N^2$ and the (differential) cross sections as $y_N^4$.
Moreover, the small-angle dominance at high $T$ makes the transport cross section
\[
\sigma_{\rm mt}\equiv \!\int\!d\Omega\,(1-\cos\theta)\,\frac{d\sigma}{d\Omega}
\]
parametrically smaller than $\sigma_{\rm el}$. The momentum-exchange (transport) rate for a heavy $N$ then reads
\begin{equation}
\gamma_p^{\rm(SM)}(N) \;\simeq\; n_{\rm bath}(T)\,\big\langle \sigma^{\rm(SM)}_{\rm mt}\,v\big\rangle\,\frac{T}{M_N}
\ \sim\  \frac{|y_N|^4\,T^2}{M_N}\,\langle 1-\cos\theta\rangle\,,
\label{eq:gammaP-N}
\end{equation}
so that, at $T\sim M_N$,
\begin{equation}
\frac{\gamma_p^{\rm(SM)}(N)}{\Gamma_D}\ \sim\ |y_N|^2\,\frac{T^2}{M_N^2}\,\langle 1-\cos\theta\rangle \ \ll\ 1
\qquad (K\gg 1)\,.
\label{eq:ratio-N}
\end{equation}
Hence one naturally realizes
\[
\underbrace{\Gamma_{\rm chem}\sim \Gamma_D\gg H}_{\text{chemical eq.}}\quad \text{while}\quad
\underbrace{\gamma_p^{\rm(SM)}(N)\ll H}_{\text{SM-kinetic decoupling}}\,,
\]
i.e.\ $N$ is chemically coupled but not fully thermalized in \emph{momentum} with the SM bath.\\

\noindent \paragraph{Momentum-space evolution and relaxation.}
The phase-space distribution $f_N(p,t)$ obeys
\begin{equation}
(\partial_t - Hp\,\partial_p) f_N \;=\; C_{\rm ID/D}[f_N;f_L,f_H] \;+\; C_{\rm el}^{\rm(SM)}[f_N;f_L,f_H]\,.
\end{equation}
Inverse decays tend to drive $f_N$ toward equilibrium, but their \emph{relaxation rate is energy dependent}:
\begin{equation}
\Gamma_{\rm relax}(E)\ \sim\ \Gamma_D\,\frac{M_N}{E}\times(\text{finite-density factors})\,,
\label{eq:relaxE}
\end{equation}
so the highest-$p$ modes equilibrate slowest (time-dilation/phase-space penalty).
When $C_{\rm el}^{\rm(SM)}$ is transport-suppressed as in \eqref{eq:gammaP-N}, the net result is:
\begin{itemize}
\item $n_N$ tracks $n_N^{\rm(eq)}$ (chemical equilibrium) for $T\!\sim\!M_N$;
\item $f_N$ can retain residual high-$p$ distortions (especially if production began non-thermally, e.g.\ from inflaton or cascade decays) because elastic refilling is too slow;
\item once inverse decays thin ($\Gamma_D/H\downarrow$), the non-thermal shape free-streams and simply redshifts (cf.\ Eq.~\eqref{eq:free-stream}).
\end{itemize}

\noindent \paragraph{Finite-density/thermal corrections.}
Thermal masses $m_H(T),m_L(T)\sim g\,T$, Pauli blocking/Bose enhancement, and off-shell scatterings modify both $C_{\rm ID/D}$ and $C_{\rm el}^{\rm(SM)}$.
Thermal field-theory treatments (HTL/resummed propagators, finite-density cutting rules) confirm that the qualitative hierarchy \eqref{eq:ratio-N} persists: washout $\propto |y_N|^2$ dominates over SM–elastic transport $\propto |y_N|^4$, with the latter further reduced by transport weighting and small-angle kinematics \cite{Beneke:2010wd}.
These effects also sharpen the energy dependence in \eqref{eq:relaxE}, reinforcing the late relaxation of high-$p$ modes.\\

\noindent \paragraph{Initial conditions and memory.}
Even in strong washout, the \emph{shape} of $f_N$ can remember non-thermal origins if they preferentially populate hard modes:
\begin{itemize}
\item \textit{Inflaton/cascade decays.} Production at $p\!\sim\!\mathcal{O}(M_\phi/2)$ (with $\phi$ heavy) seeds a hard $f_N$; inverse decays thermalize number quickly but not all momenta before $T$ drops.
\item \textit{Scattering production (UV freeze-in before equilibration).} If $N$ first appears via $2\!\to\!2$ scatterings at $T\!\gg\!M_N$, the emerging $f_N$ can be non-thermal; once $T\!\sim\!M_N$ and $K\gg1$, inverse decays pin $n_N$ while $f_N$ only partially relaxes.
\end{itemize}
Subsequent leptogenesis then proceeds with $n_N\simeq n_N^{\rm(eq)}$ but a not-quite-Maxwellian $f_N$, affecting the detailed timing of washout and CP-asymmetry build-up at the $\mathcal{O}(1)$–few$\times 10\%$ level in precision studies.\\

\noindent \paragraph{Numerical parametrics.}
Using $|y_N|\simeq \sqrt{m_\nu M_N}/v$ with $m_\nu\simeq 0.05$ eV and $v=174$ GeV gives $|y_N|\!\sim\!4\times 10^{-3}$ for $M_N=10^{10}$ GeV.
Then $\Gamma_D\!\sim\!|y_N|^2 M_N/(8\pi)\!\sim\!10^{3}$–$10^{4}$ GeV, while $H(T{=}M_N)\!\sim\!10^2$ GeV ($g_*\!\sim\!100$).
A transport estimate from \eqref{eq:gammaP-N} yields $\gamma_p^{\rm(SM)}\!\sim\!\mathcal{O}(1$–$10)$ GeV before the small-angle penalty, hence
$\Gamma_D/H\!\gg\!1$ but $\gamma_p^{\rm(SM)}/H\!\lesssim\!0.1$:
chemical equilibrium holds; full SM–kinetic equilibrium does not.\\

\noindent \paragraph{Flavor and quantum-kinetic aspects.}
At $T\!\sim\!M_N$ one may need a density-matrix (flavor) formulation for $L$ and a proper treatment of soft/collinear corrections in $C_{\rm ID/D}$.
These refinements affect the precise washout timing but not the qualitative separation $\Gamma_D\!\gg\!\gamma_p^{\rm(SM)}$ in strong washout.
Including them ensures consistent energy conservation and detailed balance in the coupled $N$–plasma system \cite{Beneke:2010wd,Giudice:2003jh}.


\subsection{Sufficient conditions for a thermal late-time shape}
Thermal velocities at late times \emph{can} occur as a result of a species being in chemical equilibrium if either:
\begin{enumerate}
\item the species remains elastically coupled to a bath with $\gamma_p/H\!\gg\!1$ well after chemical freeze-out (e.g.\ gauge-strength coupling to a relativistic plasma), or
\item the species belongs to a self-thermalized dark sector with fast internal $2\!\leftrightarrow\!2$ (and possibly $3\!\leftrightarrow\!2$) processes that maintain a common dark temperature $T_d\neq T$.
\end{enumerate}
Absent these conditions, past chemical equilibrium does not constrain the subsequent velocity distribution.


\section{Does chemical equilibrium imply kinetic equilibrium with the Standard Model plasma?}
\label{sec:chem-not-kin-with-SM}

\noindent {\bf Claim:}
No. A species can satisfy \emph{chemical} equilibrium conditions while being \emph{kinetically decoupled} from the Standard Model  bath. Chemical equilibrium constrains the abundance via number–changing reactions with rate $\Gamma_{\rm chem}\!\gg\!H$, whereas kinetic equilibrium with the SM requires a sufficiently large SM–elastic \emph{transport} rate $\gamma_p^{\rm(SM)}\!\gg\!H$ across the occupied momenta. The two are governed by different operators and kinematics (e.g.\ $s$-channel decays/annihilations vs.\ forward–peaked $t$-channel elastic scatterings), so $\Gamma_{\rm chem}$ and $\gamma_p^{\rm(SM)}$ can be parametrically and numerically separated.

\subsection*{Kinetic rationale}
Let $f(p,t)$ denote the species’ phase–space density. Its evolution obeys
\begin{equation}
(\partial_t - H\,p\,\partial_p)f \;=\; C_{\rm el}^{\rm(SM)}[f] \;+\; C_{\rm el}^{\rm(DS)}[f] \;+\; C_{\rm inel}[f]\,,
\end{equation}
where $C_{\rm inel}$ encodes number–changing reactions (to or from SM states), $C_{\rm el}^{\rm(SM)}$ elastic exchange with the SM bath, and $C_{\rm el}^{\rm(DS)}$ elastic exchange \emph{within} a dark sector (if present). Chemical equilibrium only demands $\Gamma_{\rm chem}\!\gg\!H$. Kinetic equilibrium \emph{with the SM} demands a large \emph{SM} transport rate,
\begin{equation}
\gamma_p^{\rm(SM)} \;\simeq\; n_{\rm SM}\,\big\langle\sigma_{\rm mt}^{\rm(SM)} v\big\rangle\,\frac{T}{m}\,,\qquad
\sigma_{\rm mt}\equiv\!\int d\Omega\,(1-\cos\theta)\,\frac{d\sigma}{d\Omega}\,,
\end{equation}
which is often much smaller than the \emph{inelastic} rate when elastic scattering is $t$-channel and forward–peaked (transport weighting penalizes small angles), suppressed by extra couplings, or requires heavier mediators \cite{BringmannHofmann2007,Ali-Haimoud:2018iiy}. If $\gamma_p^{\rm(SM)}\!\lesssim\!H$ while $\Gamma_{\rm chem}\!\gg\!H$, the species remains chemically tied to the SM but its momentum distribution is \emph{not} thermalized to the SM temperature. If $C_{\rm el}^{\rm(DS)}$ is fast, the species may still be thermal \emph{within a dark bath} at temperature $T_d\neq T$. We discuss two instances below.


\subsection{Conversion-driven / coannihilation freeze-out with a secluded DM state}
\label{subsec:conversion-coann}

\noindent \paragraph{Setup.}
Let $\chi$ be the DM candidate and $\psi$ a slightly heavier coannihilating partner with mass splitting
$\Delta \equiv (m_\psi-m_\chi)/m_\chi \ll 1$.
The relevant reactions are
\begin{align}
&\text{(i) conversions:} && \chi + X \;\leftrightarrow\; \psi + X' \quad\text{and}\quad \psi \;\leftrightarrow\; \chi + X \,; \label{eq:conversions}\\
&\text{(ii) (co)annihilations:} && \chi\psi,\;\psi\psi\;\;\leftrightarrow\;\; {\rm SM~SM}\,.\label{eq:coann}
\end{align}
Here $X,X'$ are SM quanta (or light dark-sector states that remain in thermal contact with the SM). The chemical relation
\begin{equation}
\mu_\chi \simeq \mu_\psi \qquad\Rightarrow\qquad \frac{n_\chi}{n_\psi} \simeq \frac{n_\chi^{\rm(eq)}}{n_\psi^{\rm(eq)}} \simeq e^{+\Delta x}\,,\qquad x\equiv \frac{m_\chi}{T}\,,
\label{eq:chemrel}
\end{equation}
is enforced by the \emph{inelastic} conversion network \eqref{eq:conversions} provided its per-particle rate
\begin{equation}
\Gamma_{\rm conv} \;\equiv\; \Gamma_{\psi\to\chi X} + n_{\rm SM}\,\langle\sigma v\rangle_{\chi X\leftrightarrow \psi X'} \;\gg\; H\,
\label{eq:convcond}
\end{equation}
throughout the epoch of interest.
The final DM abundance is then dominantly controlled by \eqref{eq:coann} (often $\psi\psi\to$ SM), i.e. by the \emph{partner’s} efficient annihilations.\\

\noindent \paragraph{Chemical vs.\ kinetic hierarchies.}
Kinetic equilibrium \emph{with the SM} requires a sufficiently large SM–elastic \emph{transport} rate for $\chi$. In many models the direct elastic portal is $t$-channel and forward–peaked (heavy mediator, small coupling), so $\sigma_{\rm mt}\ll\sigma_{\rm el}$ and $\gamma_p^{\rm(SM)}(\chi)\lesssim H$ while \emph{chemical} conversions and coannihilations still satisfy \eqref{eq:convcond}. The outcome is:
\begin{equation}
\Gamma_{\rm chem} \sim \Gamma_{\rm conv} + n_\psi\langle\sigma v\rangle_{\psi\psi} \gg H \quad \text{but} \quad \gamma_p^{\rm(SM)}(\chi) \lesssim H\,,
\label{eq:hierarchy}
\end{equation}
i.e. $\chi$ is chemically tied to the SM \emph{via} the partner network, yet kinetically decoupled from the SM bath.\\

\noindent \paragraph{Momentum-space picture.}
Conversions and inverse decays \eqref{eq:conversions} enforce \eqref{eq:chemrel} by reshuffling \emph{identities}, not by efficiently randomizing momenta. Inverse decays produce monoenergetic daughters in the parent rest-frame; $2\!\leftrightarrow\!2$ conversions typically proceed through forward–peaked $t$-channel exchange, which carries the transport penalty $(1-\cos\theta)$. Consequently, even when $n_\chi/n_\psi$ tracks \eqref{eq:chemrel}, the \emph{shape} of $f_\chi(p)$ need not be thermal at the SM temperature. Depending on the dark elastic coupling,
$\chi$ can be:
\begin{enumerate}
\item \emph{non-thermal}: if $\chi$–$\psi$ elastic scattering is also slow ($\gamma_p^{\chi\leftrightarrow\psi}\!\lesssim\!H$), $f_\chi$ develops momentum-dependent distortions during freeze-out and then free-streams (redshifts) thereafter;
\item \emph{thermalized within a dark bath}: if $\chi$–$\psi$ elastic is fast while both species are SM–decoupled, the pair attains a common \emph{dark} temperature $T_d\neq T$ (thermalized-but-secluded), even as \eqref{eq:chemrel} is maintained by inelastic conversions to SM quanta.
\end{enumerate}

\noindent \paragraph{Coupled Boltzmann system.}
A transparent way to exhibit the split is to evolve the two number densities \emph{and} at least one kinetic moment (e.g.\ a dark temperature $T_d$) or, when needed, the full $f(p,t)$. The integrated system reads schematically
\begin{align}
\dot n_\chi + 3H n_\chi &= -\langle\sigma v\rangle_{\chi\chi\to {\rm SM}} \!\left(n_\chi^2-n_{\chi,{\rm eq}}^2\right)
-\langle\sigma v\rangle_{\chi\psi\to {\rm SM}}\!\left(n_\chi n_\psi - n_{\chi,{\rm eq}} n_{\psi,{\rm eq}}\right) \nonumber\\
&\quad +\;\Gamma_{\psi\to\chi X}\!\left[n_\psi - \frac{n_{\psi,{\rm eq}}}{n_{\chi,{\rm eq}}} n_\chi\right]
+ n_{\rm SM}\langle\sigma v\rangle_{\psi X'\to\chi X}\!\left[n_\psi - \frac{n_{\psi,{\rm eq}}}{n_{\chi,{\rm eq}}} n_\chi\right], \label{eq:nchi}\\
\dot n_\psi + 3H n_\psi &= -\langle\sigma v\rangle_{\psi\psi\to {\rm SM}}\!\left(n_\psi^2-n_{\psi,{\rm eq}}^2\right)
-\langle\sigma v\rangle_{\chi\psi\to {\rm SM}}\!\left(n_\chi n_\psi - n_{\chi,{\rm eq}} n_{\psi,{\rm eq}}\right) - (\dot n_\chi + 3H n_\chi)\big|_{\rm conv}\!, \label{eq:npsi}
\end{align}
with “conv” picking the conversion terms from \eqref{eq:nchi}.
Chemical equilibrium between $\chi$ and $\psi$ corresponds to the square brackets $\propto\left[n_\psi - (n_{\psi,{\rm eq}}/n_{\chi,{\rm eq}}) n_\chi\right]$ vanishing rapidly by \eqref{eq:convcond}, irrespective of the size of the SM–elastic transport for $\chi$. Kinetic information can be tracked by a dark temperature $T_d$ obeying a heat-exchange equation (when a Maxwellian ansatz is tenable) or by a Fokker–Planck/Langevin operator for $\chi\leftrightarrow\psi$ elastic scattering; momentum-resolved treatments (full $f$) are required once high-$p$ tails or thresholds dominate the chemistry.\\

\noindent \paragraph{Parametrics.}
If the portal proceeds via a heavy mediator of mass $M$ with coupling $g_\chi g_{\rm SM}$, then
\begin{equation}
\Gamma_{\rm conv} \sim n_{\rm SM}\,\frac{(g_\chi g_{\rm SM})^2\,T^2}{M^4}\,,
\qquad
\gamma_p^{\rm(SM)}(\chi) \sim n_{\rm SM}\,\frac{(g_\chi g_{\rm SM})^2\,T^3}{M^4 m_\chi}\times \underbrace{\langle 1-\cos\theta\rangle}_{\ll 1}\!,
\label{eq:parametrics}
\end{equation}
so that $\Gamma_{\rm conv}/\gamma_p^{\rm(SM)}\sim (m_\chi/T)\langle(1-\cos\theta)^{-1}\rangle \gg 1$ for $m_\chi\!\gg\!T$ and forward–peaked kinematics. Decays $\psi\!\to\!\chi X$ add a term $\Gamma_{\psi\to\chi X}\propto g_\chi^2 \Delta m$ to \eqref{eq:convcond} that scales as $g_\chi^2$ rather than $g_\chi^4$, further widening the chemical–kinetic separation.\\

\noindent \paragraph{Phenomenological regimes.}
\begin{itemize}
\item \emph{Coannihilation with conversions fast:} $\Gamma_{\rm conv}\!\gg\!H$ and $\langle\sigma v\rangle_{\psi\psi}$ large imply standard Griest–Seckel coannihilation \cite{GriestSeckel1991}, but with the twist that $\gamma_p^{\rm(SM)}(\chi)\!\lesssim\!H$; $\chi$ is chemically tied to the SM through $\psi$, yet kinetically SM–decoupled.
\item \emph{Conversion-driven freeze-out (CDFO):} $\chi\chi$ annihilation is negligible and the relic density is set by $\chi\!\leftrightarrow\!\psi$ conversions shutting off while $\psi\psi$ annihilations remain efficient. Momentum-resolved studies show that $\chi$ can be non-thermal or thermal only within the $\{\chi,\psi\}$ subsystem at $T_d\neq T$; relic shifts at the $\mathcal{O}(10\%)$ level are common, larger when high-$p$ modes dominate the conversions \cite{GarnyHeisig2017,GarnyHeisigHufnagelLuelfVogl2019}.
\item \emph{Coscattering corridor:} when $\chi X\!\to\!\psi X'$ (with $X$ thermal) controls the abundance after $\chi\chi$ and $\psi\psi$ annihilations decouple, low-$p$ modes decouple first and $f_\chi$ becomes momentum–dependent; chemical coupling persists through \eqref{eq:conversions} even as $\gamma_p^{\rm(SM)}(\chi)\!\lesssim\!H$ \cite{DAgnoloPappadopuloRuderman2017}.
\end{itemize}

\noindent \paragraph{Representative realizations.}
Concrete examples include electroweak multiplet DM with a nearly degenerate partner where inelastic conversions proceed via weak interactions but direct $\chi$–SM elastic scattering is mass- or angle-suppressed, and simplified models with a heavy scalar or vector mediator connecting $\chi$/$\psi$ to quarks or leptons. In these cases the parametrics \eqref{eq:parametrics} obtain, and the separation in \eqref{eq:hierarchy} is realized over $x\sim 10$–$30$ during freeze-out, as explicitly demonstrated in momentum-resolved studies \cite{GarnyHeisig2017,GarnyHeisigHufnagelLuelfVogl2019}.


\subsection{\texorpdfstring{$s$}{s}-channel resonant DM annihilation with a suppressed elastic portal}
\label{subsec:s-channel-resonance}

\noindent \paragraph{Setup and kinematics.}
Let a Majorana/Dirac DM particle $\chi$ annihilate to SM states via an $s$-channel mediator $\phi$ (scalar or vector) with mass $m_\phi$ and width $\Gamma_\phi$, and couplings $g_\chi$ (to $\chi$) and $g_{\rm SM}$ (to SM). The annihilation kernel is Breit–Wigner enhanced near $s\simeq m_\phi^2$:
\begin{equation}
\sigma_{\rm ann}(s)\,v_{\rm rel}\;=\;
\frac{\mathcal{N}\,g_\chi^2 g_{\rm SM}^2\, \sqrt{1-4m_{\rm fin}^2/s}}{(s-m_\phi^2)^2 + m_\phi^2 \Gamma_\phi^2}\times \Phi(s)\,,
\label{eq:BW}
\end{equation}
with $\Phi(s)$ a smooth phase-space/initial-state factor and $\mathcal{N}$ a spin–color constant.
In the nonrelativistic regime $s\simeq 4m_\chi^2\!\left(1+\tfrac{v^2}{4}\right)$, so proximity to the pole is controlled by
\[
\delta \;\equiv\; \frac{m_\phi^2-4m_\chi^2}{4m_\chi^2}\,,
\qquad
v_{\rm res}\;\simeq\;2\sqrt{\max(\delta,0)}\,,
\qquad
\Delta v \;\sim\; \frac{\Gamma_\phi}{m_\chi}\,,
\]
i.e.\ a \emph{narrow band in velocity space} drives the largest inelastic rate. For $\delta>0$ the resonance is reachable only from the high-velocity tail; for $\delta<0$ the pole is kinematically inaccessible and annihilation proceeds off-shell but can still be sharply velocity dependent (and may receive Sommerfeld enhancement if $\phi$ mediates a long-range force).\\

\noindent \paragraph{Chemical vs.\ kinetic hierarchies.}
The annihilation (chemical) rate per particle is
\begin{equation}
\Gamma_{\rm chem}\;\equiv\; n_\chi\,\langle\sigma v\rangle_{\rm ann}
\;\simeq\; n_\chi \int\!{\rm d}^3v\, \sigma_{\rm ann}(v)\,v\, f_\chi(\vec v)\,,
\label{eq:Gammachem}
\end{equation}
which can be large when a significant fraction of $f_\chi$ overlaps the resonant band $(v\approx v_{\rm res},\,\Delta v)$.
By contrast, \emph{kinetic equilibrium with the SM} requires a large SM–elastic \emph{transport} rate $\gamma_p^{\rm(SM)}$ set by $t$-channel exchanges off relativistic targets (e.g.\ $e^\pm,\nu,\pi$). In many simplified or portal models the elastic amplitude is forward–peaked (light mediator) or contact–suppressed by a heavy propagator $M$:
\begin{equation}
\sigma^{\rm(SM)}_{\rm mt}\ \sim\
\frac{(g_\chi g_{\rm SM})^2}{(q^2+M^2)^2}\times\underbrace{\langle 1-\cos\theta\rangle}_{\ll 1\ \text{if forward-peaked}}
\ \Rightarrow\
\gamma_p^{\rm(SM)} \ \propto\ \frac{(g_\chi g_{\rm SM})^2\,T^3}{M^4\,m_\chi}\,\langle 1-\cos\theta\rangle\;.
\label{eq:mt-supp}
\end{equation}
Hence one generically finds an epoch (typically $x\equiv m_\chi/T\sim 10$–$30$) with
\begin{equation}
\Gamma_{\rm chem}\ \gg\ H\ \gtrsim\ \gamma_p^{\rm(SM)}\,,
\label{eq:ineq}
\end{equation}
i.e.\ chemical coupling is still active (because of the resonant band) while kinetic exchange with the SM fails.\\

\noindent \paragraph{Momentum-space picture and feedback.}
Because $\sigma_{\rm ann}(v)$ is sharply peaked, annihilations \emph{deplete a narrow shell in momentum space} around $p_{\rm res}\simeq m_\chi v_{\rm res}$.
If $\gamma_p^{\rm(SM)}\ll H$, elastic scattering on the SM cannot efficiently refill this shell; $f_\chi$ develops a “notch” or dip, and the system departs from a single-temperature form. Two robust consequences follow:
\begin{enumerate}
\item \emph{Thermal-averaging bias.} The common kinetic-equilibrium estimate $\langle\sigma v\rangle_{\rm ann}^{\rm(KE)}\!=\!\langle\sigma v\rangle$ evaluated on a Maxwellian at $T_\chi\!=\!T$ (or at a fitted $T_\chi$) \emph{mis-estimates} the true average in \eqref{eq:Gammachem} once the resonant band has been sculpted out of $f_\chi$. This shifts the freeze-out condition and the relic density—by factors from $\mathcal{O}(1)$ up to an order of magnitude in sharp/narrow cases—relative to kinetic-equilibrium treatments \cite{Binder:2017rgn,Ala-Mattinen:2022nnh,Aboubrahim:2023pyr}.
\item \emph{Velocity-history memory.} The post-freeze-out $f_\chi$ retains a memory of the resonant depletion. Late-time observables that depend on $v$ (indirect detection, CMB energy injection) can therefore differ from predictions made with a Maxwellian $f_\chi$ at the same number density.
\end{enumerate}

\noindent \paragraph{Representative realizations.}
\begin{itemize}
\item \emph{Higgs-portal resonance:} $m_\chi\simeq m_h/2$ with annihilation via $h^\ast\to b\bar b,\tau^+\tau^-$. The chemical rate is resonantly enhanced near freeze-out, whereas SM–elastic scattering on relativistic leptons via $t$-channel $h$ exchange is Yukawa-suppressed ($\propto y_\ell^2$) and contact-like ($\propto m_h^{-4}$), driving $\gamma_p^{\rm(SM)}\!\lesssim\!H$ even while $\Gamma_{\rm chem}\!\gg\!H$.
\item \emph{$Z'$ near threshold:} a narrow vector mediator with $m_{Z'}\simeq 2m_\chi$ and small kinetic mixing $\epsilon$ can produce large $\Gamma_{\rm chem}$ while $\chi$–$e^\pm$ elastic scattering is both $\epsilon^2$- and transport-suppressed, realizing \eqref{eq:ineq}.
\end{itemize}

\noindent \paragraph{Sommerfeld and bound-state effects.}
If the mediator participates in the initial-state potential, Sommerfeld factors $S(v)$ can further enhance annihilation at small $v$ (or generate resonant peaks when a near-threshold bound state forms). This \emph{sharpens} the momentum selectivity and strengthens the need for momentum-resolved treatments; $S(v)$ multiplies the Breit–Wigner in \eqref{eq:BW} and accentuates the depletion of specific velocity shells. Kinetic failure with the SM then occurs even earlier because annihilations preferentially target the lowest-$v$ modes while elastic refilling remains slow.\\

\noindent \paragraph{Coupled evolution: from full $f$ to controlled closures.}
A careful treatment evolves the phase-space density $f_\chi(p,t)$ with the exact $2\!\leftrightarrow\!2$ annihilation kernel and an elastic operator appropriate to the scattering regime:
\begin{equation}
(\partial_t - Hp\partial_p)f_\chi \;=\; C_{\rm ann}[f_\chi] \;+\; C_{\rm el}^{\rm(SM)}[f_\chi]\,.
\end{equation}
In forward–peaked regimes a matched Fokker–Planck/Langevin operator for $C_{\rm el}^{\rm(SM)}$ (drag/diffusion chosen to reproduce momentum and heat exchange) yields reliable \emph{transport} rates and captures the failure of refilling the resonant band. Moment closures (number+kinetic temperature $T_\chi$) can work if $\sigma_{\rm ann}(v)$ is broad; they underperform when the resonance is narrow or when thresholds select the high-$p$ tail \cite{Binder:2017rgn,Ala-Mattinen:2022nnh,Aboubrahim:2023pyr}.\\

\noindent \paragraph{Parametrics and scaling.}
For a heavy $t$-channel mediator of mass $M$,
\[
\gamma_p^{\rm(SM)} \sim \frac{(g_\chi g_{\rm SM})^2}{M^4}\,\frac{T^3}{m_\chi}\times\langle 1-\cos\theta\rangle\,,
\qquad
\Gamma_{\rm chem} \sim n_\chi\,\frac{g_\chi^2 g_{\rm SM}^2}{(4m_\chi^2-m_\phi^2)^2+m_\phi^2\Gamma_\phi^2}\,,
\]
so that
$\Gamma_{\rm chem}/\gamma_p^{\rm(SM)} \propto (n_\chi m_\chi/T^3)\,[\langle 1-\cos\theta\rangle]^{-1}\,[(4m_\chi^2-m_\phi^2)^2+m_\phi^2\Gamma_\phi^2]^{-1}$.
At $x\!\sim\!20$ this ratio is naturally large unless elastic scattering is gauge-strength and unsuppressed by transport weighting.\\

In sum, near-threshold $s$-channel annihilation furnishes a concrete regime where \emph{chemical} coupling is strong—thanks to a narrow velocity band—while \emph{kinetic} coupling to the SM bath is weak due to transport-suppressed elastic scattering. Momentum-resolved studies consistently find pronounced spectral features and sizable relic-density shifts relative to kinetic-equilibrium treatments \cite{GriestSeckel1991,Binder:2017rgn,Ala-Mattinen:2022nnh,Aboubrahim:2023pyr}.


\subsection{Sufficient conditions for kinetic equilibrium \emph{with the SM} for species in chemical equilibrium}
Kinetic equilibrium with the SM \emph{does} follow from chemical equilibrium if either:
\begin{enumerate}
\item the elastic portal to SM gauge-charged targets is strong enough that $\gamma_p^{\rm(SM)}/H\!\gg\!1$ throughout and after the chemically active epoch (typical for species with unsuppressed gauge interactions), or
\item the inelastic channel that enforces chemistry \emph{also} implies large-angle elastic scattering on SM targets with comparable rate and no transport suppression.
\end{enumerate}
Absent these conditions, chemical equilibrium does not imply kinetic equilibrium with the SM.


\section{Does chemical equilibrium imply kinetic equilibrium within a dark sector?}
\label{sec:chem-vs-kin-DS}

\noindent {\bf Claim:}
No. In a dark sector (DS) with its own temperature $T_d$, \emph{chemical} equilibrium among dark species does not, in general, guarantee \emph{kinetic} equilibrium (i.e.\ thermal velocity distributions) even solely within the DS. Chemical balance fixes chemical potentials through the reaction network, $\sum_j \nu_j \mu_j=0$ for each active process $\sum_j \nu_j X_j\leftrightarrow 0$, provided the slowest number–changing rate $\Gamma_{\rm chem}\gg H$. By contrast, kinetic equilibrium requires sufficiently rapid \emph{momentum exchange} inside the DS, quantified by transport rates
\begin{equation}
\gamma_p^{\rm(DS)}(i)\;\simeq\; \sum_{t\in{\rm DS}} n_t(T_d)\,\big\langle\sigma^{\rm(DS)}_{{\rm mt},\,it} v\big\rangle\,\frac{T_d}{m_i}\ \gg\ H.
\label{eq:gammaDS}
\end{equation}
Because chemical and transport operators scale differently with couplings, mediator masses and kinematics (e.g.\ $s$-channel decays/annihilations vs.\ forward–peaked $t$-channel elastic), one can have $\Gamma_{\rm chem}\!\gg\!H$ while $\gamma_p^{\rm(DS)}\!\lesssim\!H$ for part of the populated momentum range.

\subsection*{Kinetic rationale within the DS}
Let $f_i(p,t)$ be the phase–space density for dark species $i$. Its evolution obeys
\begin{equation}
(\partial_t-Hp\,\partial_p) f_i \;=\; C_{\rm el}^{\rm(DS)}[f] \;+\; C_{\rm inel}^{\rm(DS)}[f]\ (+\ C_{\rm portal}[f]\ \text{if present})\,.
\label{eq:Boltzmann-DS}
\end{equation}
When $C_{\rm inel}^{\rm(DS)}$ is fast it enforces chemical relations (shared $\mu_i$’s or their linear constraints), but the \emph{shape} of $f_i$ thermalizes only if $C_{\rm el}^{\rm(DS)}$ provides rapid, transport–efficient momentum exchange across momenta carrying $n$ and $\rho$. If $\gamma_p^{\rm(DS)}\!\lesssim\!H$ after some $t_{\rm kd}^{\rm(DS)}$, the subsequent spectrum is a redshifted memory of $f_i$ at kinetic decoupling plus any later injections, and need not be Maxwell–Boltzmann/Fermi–Dirac at $T_d$.\\


\subsection{Two-state DS with conversions but slow DS--elastic}
\label{subsec:DS-conversion}

\noindent \paragraph{Field content and reactions.}
Consider a secluded dark sector (DS) with temperature $T_d\neq T$, containing a DM state $\chi$ and a slightly heavier partner $\psi$ with
$\Delta\equiv (m_\psi-m_\chi)/m_\chi\ll 1$, plus light dark radiation $\varphi$ that efficiently self-thermalizes and sets $T_d$.
The key processes are
\begin{align}
&\text{(inelastic conversions)} &&
\chi \leftrightarrow \psi+\varphi,\qquad
\chi\,\varphi \leftrightarrow \psi\,\varphi', \label{eq:DS-conv}\\
&\text{(dark elastic)} &&
\chi\,\varphi \leftrightarrow \chi\,\varphi,\quad
\chi\,\psi \leftrightarrow \chi\,\psi,\quad
\psi\,\varphi \leftrightarrow \psi\,\varphi, \label{eq:DS-elastic}\\
&\text{(number depletion)} &&
\psi\psi \leftrightarrow \text{light DS (or SM)}. \label{eq:DS-ann}
\end{align}
We assume $\varphi$ is light enough (or gauge-like) that \eqref{eq:DS-elastic} keeps $\varphi$ internally thermal with negligible chemical potential.\\

\noindent \paragraph{Chemical equilibrium at $T_d$ vs.\ kinetic equilibrium within the DS.}
The inelastic network \eqref{eq:DS-conv} enforces chemical relations
\begin{equation}
\mu_\chi\simeq \mu_\psi \qquad \Rightarrow \qquad
\frac{n_\chi}{n_\psi}\simeq \frac{n_\chi^{\rm(eq)}(T_d)}{n_\psi^{\rm(eq)}(T_d)} \simeq e^{+\Delta x_d}\,,
\qquad x_d\equiv \frac{m_\chi}{T_d},
\label{eq:DS-chem}
\end{equation}
provided the slowest per-particle conversion rate satisfies
\begin{equation}
\Gamma_{\rm conv}^{\rm(DS)} \;\equiv\;
\Gamma_{\psi\to\chi\varphi}
+ n_\varphi\langle \sigma v\rangle_{\chi\varphi\leftrightarrow\psi\varphi'}
\gg H. \label{eq:DS-Gamma-conv}
\end{equation}
By contrast, \emph{kinetic} equilibrium of $\chi$ \emph{within the DS} requires a large \emph{transport} rate from \eqref{eq:DS-elastic}:
\begin{equation}
\gamma_p^{\rm(DS)}(\chi)
\simeq \sum_{t=\{\varphi,\psi\}} n_t(T_d)\,
\big\langle \sigma_{{\rm mt},\,\chi t}^{\rm(DS)} v\big\rangle\,
\frac{T_d}{m_\chi}\ \gg\ H,
\qquad
\sigma_{\rm mt}\equiv\int d\Omega\,(1-\cos\theta)\,\frac{d\sigma}{d\Omega}, \label{eq:DS-transport}
\end{equation}
with the $(1-\cos\theta)$ weighting penalizing forward scattering.
A broad and generic hierarchy then emerges:
\begin{equation}
\Gamma_{\rm chem}^{\rm(DS)}\sim \Gamma_{\rm conv}^{\rm(DS)} \gg H
\quad \text{while} \quad
\gamma_p^{\rm(DS)}(\chi)\lesssim H,
\label{eq:DS-hierarchy}
\end{equation}
whenever elastic exchange is dominated by $t$-channel exchange of a heavier mediator or suffers additional coupling/angle suppressions.\\

\noindent \paragraph{Parametrics from a simple mediator model.}
Take conversions from a Yukawa-like interaction $y\,\psi\chi\varphi$ and DS elastic mediated by a heavy vector/scalar of mass $M_d$ with coupling $g_d$.
For $\Delta\ll 1$ (small $Q$-value two-body decay) one finds
\begin{equation}
\Gamma_{\psi\to\chi\varphi}\ \sim\ \frac{y^2}{16\pi}\,\Delta\,m_\chi,
\qquad
n_\varphi\langle \sigma v\rangle_{\chi\varphi\to\psi\varphi'} \ \sim\ y^2\,T_d,
\label{eq:DS-conv-param}
\end{equation}
while the \emph{transport} rate from $\chi$--$\varphi$ elastic scattering scales as
\begin{equation}
\gamma_p^{\rm(DS)}(\chi)\ \sim\
n_\varphi\,\frac{g_d^4\,T_d^2}{M_d^4}\,\frac{T_d}{m_\chi}\times\langle 1-\cos\theta\rangle.
\label{eq:DS-gammap-param}
\end{equation}
Hence
\[
\frac{\Gamma_{\rm conv}^{\rm(DS)}}{\gamma_p^{\rm(DS)}(\chi)}\ \sim\
\left(\frac{y^2}{g_d^4}\right)\left(\frac{M_d^4}{T_d^2}\right)\left(\frac{m_\chi}{T_d}\right)\,\langle 1-\cos\theta\rangle^{-1},
\]
which is generically $\gg 1$ for $m_\chi\!\gg\!T_d$ and forward-peaked elastic exchange (small $\langle 1-\cos\theta\rangle$), realizing \eqref{eq:DS-hierarchy} over the freeze-out window $x_d\sim 10$--$30$.\\

\noindent \paragraph{Momentum-space picture.}
Inverse decays $\chi\varphi\!\to\!\psi$ and decays $\psi\!\to\!\chi\varphi$ reshuffle \emph{identities} and inject quanta at characteristic momenta set by the $Q$-value ($\sim \Delta m_\chi$).
When $\gamma_p^{\rm(DS)}(\chi)\!\lesssim\!H$, elastic refilling is inefficient and $f_\chi(p)$ develops non-thermal features:
(i) a hard tail from decay-produced $\chi$,
(ii) momentum-selective dips/excesses from the $2\!\leftrightarrow\!2$ conversion kernel.
Even while \eqref{eq:DS-chem} holds at the level of number ratios, the velocity distribution of $\chi$ need not be Maxwell–Boltzmann at $T_d$.

\noindent \paragraph{Coupled evolution (number, heat, and, when needed, full $f$).}
A minimal yet informative closure supplements the number densities with a separate kinetic temperature for the heavy sector, $T_\chi$, in addition to the DS bath temperature $T_d$ (carried by $\varphi$):
\begin{align}
\dot n_\chi +3H n_\chi &= -\mathcal{C}_{\rm dep}[n_\chi,n_\psi;T_d]\;+\;\mathcal{C}_{\rm conv}[n_\chi,n_\psi;T_d], \label{eq:DS-nchi}\\
\frac{3}{2}\Big(\dot n_\chi\,T_\chi + n_\chi\,\dot T_\chi\Big)+5H n_\chi T_\chi
&= -\underbrace{\mathcal{Q}_{\rm el}}_{\propto\gamma_p^{\rm(DS)}(\chi)\,(T_\chi-T_d)}
\;+\; \underbrace{\mathcal{Q}_{\rm conv}}_{\text{energy~injection/extraction from conversions}}
\;+\; \ldots \label{eq:DS-heat}
\end{align}
Here $\mathcal{C}_{\rm dep}$ lumps number-depleting reactions (often dominated by $\psi\psi\!\to$ light DS/SM), while $\mathcal{C}_{\rm conv}$ enforces \eqref{eq:DS-chem} when \eqref{eq:DS-Gamma-conv} holds. The elastic heat-exchange term $\mathcal{Q}_{\rm el}$ drives $T_\chi\!\to\!T_d$ with rate $\sim\gamma_p^{\rm(DS)}$, whereas $\mathcal{Q}_{\rm conv}$ accounts for the non-thermal energy transfer carried by decay/ inverse-decay quanta. This two-temperature (cBE) system is accurate if $f_\chi$ remains close to Maxwellian; otherwise a momentum-resolved treatment,
\[
(\partial_t - Hp\partial_p)f_\chi = C_{\rm conv}[f_\chi,f_\psi,f_\varphi]\;+\;C_{\rm el}^{\rm(DS)}[f_\chi,f_\varphi],
\]
is required. In small-angle regimes $C_{\rm el}^{\rm(DS)}$ can be implemented as a matched Fokker–Planck/Langevin operator that reproduces the exact momentum/heat exchange and yields $\gamma_p^{\rm(DS)}$ directly.\\

\noindent \paragraph{Regimes and outcomes.}
\begin{itemize}
\item \emph{Non-thermal $\chi$:} if both $\chi$--$\varphi$ and $\chi$--$\psi$ elastic are slow, \eqref{eq:DS-heat} gives $T_\chi/T_d \to$ constant $\gtrsim 1$ (or a slowly varying ratio) and $f_\chi$ develops persistent high-$p$ features set by conversion kinematics. The relic density can shift relative to a single-$T_d$ closure when the chemistry is sensitive to high-$p$ modes.
\item \emph{Thermalized-but-secluded pair $\{\psi,\varphi\}$:} if $\psi$--$\varphi$ elastic is fast while $\chi$--$\varphi$ is slow, the bath and $\psi$ share $T_d$ and Maxwellian shapes; $\chi$ tracks \eqref{eq:DS-chem} number-wise but remains kinetically misaligned (non-thermal or at a distinct $T_\chi$).
\item \emph{Recovery of DS kinetic equilibrium:} if at later times a lighter mediator turns on (phase transition) or the forward peak is screened (finite-$T_d$ mass), $\gamma_p^{\rm(DS)}$ can exceed $H$ again, partially thermalizing $\chi$ at late $x_d$; imprints of earlier non-thermality may nevertheless survive in the low-$v$ tail.
\end{itemize}

\noindent \paragraph{Representative realizations.}
(i) A dark-$U(1)$ with heavy dark photon $A'$ mediating elastic scattering ($g_d$) and a Yukawa $y\,\psi\chi\varphi$ driving conversions; (ii) $Z_2$-odd doublet of fermions with a light pseudoscalar $\varphi$ enabling $\psi\!\to\!\chi\varphi$ while elastic $\chi$--$\varphi$ proceeds via heavy scalar exchange; (iii) multi-TeV $\chi$ with $\Delta\!\sim\!0.05$ and $M_d\!\gg\!m_\chi$, realizing \eqref{eq:DS-hierarchy} for $x_d\!\sim\!10$--$30$.\\

Note that this DS example is the secluded analogue of conversion/coannihilation and coscattering dynamics discussed for SM-coupled sectors (see \S\ref{subsec:conversion-coann}); the kinetic diagnostics (transport-weighted rates, FP/Langevin) follow the methodology developed for early kinetic decoupling and momentum-resolved freeze-out \cite{BringmannHofmann2007,Ali-Haimoud:2018iiy}. Dark-sector temperature evolution and cannibal phases provide additional structure when $3\!\to\!2$ is active \cite{PappadopuloRuderman2016,FarinaPappadopuloRudermanTrevisan2016}; conversion-driven freeze-out and related multi-state dynamics are discussed in \cite{GarnyHeisig2017,GarnyHeisigHufnagelLuelfVogl2019}.


\subsection{Semi-annihilation, $Z_3$-like DS: chemistry without full kineticization}
\label{subsec:semiann}

\noindent \paragraph{Field content and reactions.}
Consider a secluded dark sector (DS) with temperature $T_d\neq T$, containing a stable DM particle $\chi$ (stabilized by a $Z_3$-like symmetry), a light dark state $\phi$ that sets $T_d$ via fast self-interactions, and semi-annihilation (SA)
\begin{equation}
\chi\,\chi \;\leftrightarrow\; \chi\,\phi\,,
\label{eq:SA}
\end{equation}
supplemented by rapid $\phi$-self-thermalization (e.g.\ $\phi\phi\leftrightarrow\phi\phi$ and, optionally, $\phi\phi\leftrightarrow\phi\phi\phi$ or $\phi\to$ light DS) and \emph{suppressed} elastic $\chi$–$\phi$ scattering (heavy $t$-channel mediator or forward-peaked kinematics). Equation~\eqref{eq:SA} changes the $\chi$ number by $-1$ per reaction and links the chemical potentials,
\begin{equation}
2\mu_\chi = \mu_\chi + \mu_\phi \quad\Rightarrow\quad \mu_\chi = \mu_\phi\,,
\label{eq:chem-SA}
\end{equation}
so that, in the presence of fast SA and any $\phi$-number–violating processes (e.g.\ $\phi$ decay or $3\!\leftrightarrow\!2$ in the $\phi$-sector), one obtains $\mu_\phi\simeq 0$ and thus $\mu_\chi\simeq 0$. If $\phi$ only undergoes $2\!\leftrightarrow\!2$ self-scattering, its number is conserved and the common value $\mu_\chi=\mu_\phi$ can be nonzero; in either case, \emph{chemical} balance is enforced at the DS temperature $T_d$ when the slowest number-changing rate satisfies $\Gamma_{\rm chem}^{\rm(DS)}\gg H$.\\

\noindent \paragraph{Rate hierarchy and transport.}
Define the SA depletion rate per $\chi$ as
\begin{equation}
\Gamma_{\rm SA} \;\equiv\; n_\chi\,\big\langle \sigma v\big\rangle_{\chi\chi\to\chi\phi}(T_d,T_\chi)\,,
\qquad
\Gamma_{\rm chem}^{\rm(DS)}\sim \Gamma_{\rm SA}\,,
\label{eq:GammaSA}
\end{equation}
where the thermal average may depend on both $T_d$ (for $\phi$) and the \emph{actual} $\chi$ spectrum (see below).
Kinetic equilibrium \emph{within the DS} requires a large \emph{transport} rate for momentum exchange between $\chi$ and the bath,
\begin{equation}
\gamma_p^{\rm(DS)}(\chi)\;\simeq\; n_\phi(T_d)\,\big\langle \sigma_{\rm mt}(\chi\phi\to\chi\phi)\, v\big\rangle\,\frac{T_d}{m_\chi}\,,
\qquad
\sigma_{\rm mt}\equiv \!\int d\Omega\,(1-\cos\theta)\,\frac{d\sigma}{d\Omega}\,.
\label{eq:gamma-chi-phi}
\end{equation}
With heavy $t$-channel mediators or light mediators producing forward peaking, $\sigma_{\rm mt}\ll\sigma_{\rm el}$ and thus $\gamma_p^{\rm(DS)}(\chi)\lesssim H$ over $x_d\equiv m_\chi/T_d\sim 10$–$30$ even when $\Gamma_{\rm SA}\gg H$.
Therefore one naturally realizes
\begin{equation}
\Gamma_{\rm chem}^{\rm(DS)} \gg H \quad\text{while}\quad \gamma_p^{\rm(DS)}(\chi) \lesssim H\,,
\label{eq:hier-SA}
\end{equation}
i.e.\ chemical equilibrium in the DS without full kineticization of $\chi$ at $T_d$.\\

\noindent \paragraph{Momentum-space picture: injection vs.\ refilling.}
Semi-annihilation is \emph{momentum-selective}. In the nonrelativistic regime, two nearly at-rest $\chi$’s produce a final-state $\chi$ and a (typically relativistic) $\phi$. The outgoing $\chi$ receives a kinetic kick set by the $Q$-value:
\[
Q \simeq m_\chi - m_\phi + \mathcal{O}(T_d) \ \approx\ m_\chi \quad (\text{if } m_\phi\ll m_\chi)\,,
\]
so SA \emph{injects} a population of hard $\chi$ at $p\sim \sqrt{2 m_\chi Q}$, while inverse SA, $\chi\phi\to\chi\chi$, repopulates specific momentum bands governed by the $\phi$ bath. When $\gamma_p^{\rm(DS)}(\chi)\!\lesssim\!H$, elastic $\chi$–$\phi$ scattering cannot efficiently isotropize/thermalize these injections. The result is a non-thermal $f_\chi(p)$ featuring (i) a high-$p$ tail from $\chi\chi\to\chi\phi$ and (ii) momentum-band features from $\chi\phi\to\chi\chi$; both features simply redshift once DS kinetic decoupling is complete.\\

\noindent \paragraph{Energetics and dark self-heating.}
Each SA event converts approximately one rest mass $m_\chi$ into kinetic energy shared by $\chi$ and $\phi$. In a two-temperature closure the heavy component obeys
\begin{equation}
\frac{3}{2}\Big(\dot n_\chi T_\chi + n_\chi \dot T_\chi\Big)+5H n_\chi T_\chi
\;=\; -\underbrace{\mathcal{Q}_{\rm el}}_{\propto \gamma_p^{\rm(DS)}(\chi)\,(T_\chi-T_d)}
\;+\; \underbrace{\mathcal{Q}_{\rm SA}}_{\simeq\,\tfrac{1}{2} m_\chi\,\Gamma_{\rm SA}\,n_\chi}
\;+\; \ldots
\label{eq:heat-SA}
\end{equation}
where $\mathcal{Q}_{\rm el}$ thermalizes $T_\chi\to T_d$ at a rate set by \eqref{eq:gamma-chi-phi}, while $\mathcal{Q}_{\rm SA}>0$ \emph{heats} the heavy component (the $\tfrac{1}{2}$ reflects that one $\chi$ is removed per reaction). If $\gamma_p^{\rm(DS)}\!\ll\!\Gamma_{\rm SA}$ during freeze-out, semi-annihilation drives $T_\chi/T_d$ above unity (``partial cannibalization by SA'') and accentuates non-thermal high-$p$ features; the bath $\phi$ also receives heat, but if $g_{*,d}$ is large its temperature responds only mildly.

\noindent \paragraph{Coupled number evolution with detailed balance.}
The number density evolves as
\begin{equation}
\dot n_\chi + 3H n_\chi \;=\; -\,\langle \sigma v\rangle_{\chi\chi\to\chi\phi}[f_\chi,f_\phi]\; n_\chi^2
\;+\;
\langle \sigma v\rangle_{\chi\phi\to\chi\chi}[f_\chi,f_\phi]\; n_\chi n_\phi\,,
\label{eq:n-SA}
\end{equation}
with \emph{distinct} thermal averages for forward and inverse processes once $f_\chi$ is non-thermal or $T_\chi\neq T_d$. Detailed balance at the microphysical level still relates the \emph{kernels}, but the phase-space–averaged rates differ from those computed assuming a Maxwellian at $T_d$. In practice:
\begin{itemize}
\item A single-temperature closure ($T_\chi\!=\!T_d$) \emph{over-}mixes momenta and misestimates the forward/inverse balance when injection shapes are important.
\item A two-temperature (cBE) system $(n_\chi, T_\chi; T_d)$ captures leading effects if $f_\chi$ remains nearly Maxwellian; otherwise a momentum-resolved treatment is required.
\end{itemize}

\noindent \paragraph{Parametric example.}
Take a $Z_3$-like interaction $\lambda\,\chi^3\phi + \text{h.c.}$ generating SA at
\[
\langle\sigma v\rangle_{\rm SA} \sim \frac{|\lambda|^2}{16\pi\,m_\chi^2}\,,
\]
and elastic $\chi$–$\phi$ scattering mediated by a heavy state of mass $M_d$ with coupling $g_d$:
\[
\gamma_p^{\rm(DS)}(\chi) \ \sim\ n_\phi\,\frac{g_d^4\,T_d^2}{M_d^4}\,\frac{T_d}{m_\chi}\times \langle 1-\cos\theta\rangle\,.
\]
For $m_\chi\!\gg\!T_d$ and forward peaking, $\Gamma_{\rm SA}/\gamma_p^{\rm(DS)} \propto (m_\chi/T_d)\,(M_d^4/g_d^4)\,\langle 1-\cos\theta\rangle^{-1}\gg 1$, realizing \eqref{eq:hier-SA} throughout freeze-out.\\

\noindent \paragraph{Regimes and outcomes.}
\begin{itemize}
\item \emph{Chemistry fast, DS–elastic slow:} $n_\chi$ tracks its chemical trajectory at $T_d$ (set by \eqref{eq:chem-SA}) while $f_\chi$ exhibits non-thermal high-$p$ shoulders and $T_\chi>T_d$. Relic-density predictions using $T_\chi\!=\!T_d$ can err at the $\mathcal{O}(10\%)$–$\mathcal{O}(1)$ level when SA is sharply momentum-selective.
\item \emph{Late partial re-kineticization:} If $\gamma_p^{\rm(DS)}$ grows relative to $H$ (e.g.\ screening of the forward peak, mediator becoming lighter, or $n_\phi$ rising during a DS phase transition), the hard tail is partially erased at $x_d\!\gtrsim\!x_{\rm fo}$; observables sensitive to the low-$v$ tail (indirect/CMB) can still retain memory of early non-thermality.
\end{itemize}

Semi-annihilation provides a clean intra-DS realization of “chemistry without kineticization”: number-changing is efficient, but DS–elastic transport can be parametrically weaker. The energetics mimic cannibal self-heating to be discussed below (\S\ref{subsec:cannibal-caveat}) but proceed through $2\!\leftrightarrow\!2$ reactions that \emph{inject} hard $\chi$ rather than preferentially depleting low-$p$ modes. For model scaffolding and phenomenology of semi-annihilation in $Z_3$-like sectors, see e.g.\ \cite{DEramo:2010keq}; kinetic diagnostics and transport-weighted treatments follow \cite{BringmannHofmann2007,Ali-Haimoud:2018iiy}.


\subsection{Cannibal sector caveat: $3\!\to\!2$ chemistry with marginal $2\!\to\!2$}
\label{subsec:cannibal-caveat}

\noindent \paragraph{Field content and standard assumption.}
Consider a secluded dark sector (DS) at temperature $T_d\neq T$ with a single nonrelativistic species $\chi$ undergoing efficient number-changing
\begin{equation}
\chi\chi\chi \;\leftrightarrow\; \chi\chi\,,
\label{eq:three-two}
\end{equation}
and (possibly) elastic self-scattering $\chi\chi\!\leftrightarrow\!\chi\chi$.
In the usual \emph{cannibal} regime one assumes fast $2\!\to\!2$ elastic so that $f_\chi$ remains Maxwell–Boltzmann at a common $T_d$, while \eqref{eq:three-two} maintains chemical equilibrium and heats the sector, yielding the familiar $T_d\propto 1/\ln a$ during the nonrelativistic phase \cite{PappadopuloRuderman2016,FarinaPappadopuloRudermanTrevisan2016}.\\

\noindent \paragraph{Chemical vs.\ kinetic hierarchy inside the DS.}
Let the slowest number–changing per-particle rate be
\begin{equation}
\Gamma^{\rm(DS)}_{3\to2} \;\equiv\; n_\chi^2\,\big\langle \sigma v^2\big\rangle_{3\to 2}(T_d)\,,
\label{eq:gamma32}
\end{equation}
which can readily satisfy $\Gamma^{\rm(DS)}_{3\to2}\gg H$ around freeze-out.
Kinetic equilibrium \emph{within the DS} requires a large \emph{transport} rate from elastic self-scattering,
\begin{equation}
\gamma_p^{\rm(DS)}(\chi)\;\simeq\; n_\chi\,\big\langle\sigma_{\rm mt}(\chi\chi\to\chi\chi)\, v\big\rangle\,\frac{T_d}{m_\chi}\,,
\qquad
\sigma_{\rm mt}\equiv\!\int d\Omega\,(1-\cos\theta)\,\frac{d\sigma}{d\Omega}\,,
\label{eq:gamma-el}
\end{equation}
which is suppressed if the dominant $2\!\to\!2$ topology is forward–peaked $t$-channel or proceeds via a heavy mediator.
When
\begin{equation}
\Gamma^{\rm(DS)}_{3\to2} \gg H \quad \text{but} \quad \gamma_p^{\rm(DS)}(\chi)\lesssim H\,,
\label{eq:cannibal-split}
\end{equation}
the DS maintains chemical relations and number depletion (``chemistry'') while failing to fully isotropize/thermalize momenta (``kinetics'') on a Hubble time.\\

\noindent \paragraph{Momentum-space picture.}
In the nonrelativistic regime, the $3\!\to\!2$ process is exothermic: two survivors share the rest-mass energy of the third.
If elastic transport is slow, repeated $3\!\to\!2$ events deplete low-$p$ quanta and \emph{inject} kinetic energy into the surviving pair, pushing weight toward intermediate/high $p$.
Absent rapid $\chi\chi\!\leftrightarrow\!\chi\chi$ to redistribute momenta,
\begin{equation}
(\partial_t - Hp\partial_p)f_\chi \;=\; C_{3\leftrightarrow 2}[f_\chi]\;+\; C^{\rm(DS)}_{\rm el}[f_\chi],
\qquad \gamma_p^{\rm(DS)}\ll H \;\Rightarrow\; f_\chi(p,t)\ \text{non-thermal}\,,
\end{equation}
with characteristic signatures:
(i) selective depletion at low $p$ (where $3\!\to\!2$ is most phase-space efficient in the NR limit),
(ii) a self-heated shoulder at higher $p$ from the exothermic recoil.
These features then \emph{redshift} after intra-DS kinetic decoupling.\\

\noindent \paragraph{Energetics and temperature evolution beyond the textbook limit.}
Write the number and heat equations (integrated Boltzmann hierarchy) for a heavy NR component:
\begin{align}
\dot n_\chi + 3H n_\chi &= -\,\big\langle\sigma v^2\big\rangle_{3\to 2}\,\Big(n_\chi^3 - n_\chi^2 n_{\chi,{\rm eq}}(T_d)\Big)\,, \label{eq:n-32}\\
\frac{3}{2}\Big(\dot n_\chi T_\chi + n_\chi \dot T_\chi\Big)+5H n_\chi T_\chi
&= -\,\underbrace{\mathcal{Q}_{\rm el}}_{\propto\,\gamma_p^{\rm(DS)}(\chi)\,(T_\chi-T_d)}
\;+\;\underbrace{\mathcal{Q}_{3\to2}}_{\text{self-heating from }3\!\to\!2}
\;+\;\ldots\label{eq:heat-32}
\end{align}
The standard cannibal result $T_\chi=T_d\propto 1/\ln a$ follows when $\mathcal{Q}_{\rm el}$ enforces $T_\chi\!=\!T_d$ at all times.
When $\gamma_p^{\rm(DS)}\!\lesssim\!H$, however, $\mathcal{Q}_{\rm el}$ cannot enforce $T_\chi\to T_d$ and \eqref{eq:heat-32} yields $T_\chi/T_d>1$ with a slower-than-logarithmic cooling of the \emph{heavy} component, reflecting incomplete kineticization of the self-heating.
This modifies both the instantaneous $\langle\sigma v^2\rangle_{3\to2}$ (through its temperature/momentum dependence) and the mapping between comoving entropy conservation in the DS and the $T_d(a)$ trajectory familiar from \cite{PappadopuloRuderman2016,FarinaPappadopuloRudermanTrevisan2016}.\\

\noindent \paragraph{Parametrics and when the split occurs.}
Assume $3\!\to\!2$ arises from a contact operator with strength $\alpha_d/m_\chi^5$ (NR scaling), while elastic proceeds via a heavy mediator of mass $M_d$ and coupling $g_d$:
\begin{equation}
\Gamma^{\rm(DS)}_{3\to2} \sim \alpha_d\,\frac{n_\chi^2}{m_\chi^5}\,,
\qquad
\gamma_p^{\rm(DS)}(\chi) \sim n_\chi\,\frac{g_d^4\,T_d^2}{M_d^4}\,\frac{T_d}{m_\chi}\times \langle 1-\cos\theta\rangle\,.
\end{equation}
At $x_d\equiv m_\chi/T_d\simeq 10$–$30$ one then finds
\[
\frac{\Gamma^{\rm(DS)}_{3\to2}}{\gamma_p^{\rm(DS)}(\chi)}
\ \propto\
\left(\frac{\alpha_d}{g_d^4}\right)\left(\frac{M_d^4}{m_\chi^4}\right)\left(\frac{m_\chi}{T_d}\right)\,\langle 1-\cos\theta\rangle^{-1}
\ \gg\ 1
\]
for modest couplings and forward–peaked elastic, realizing \eqref{eq:cannibal-split} over the epoch where $3\!\to\!2$ controls the chemistry.\\

\noindent \paragraph{When \texorpdfstring{$3\!\to\!2$}{3→2} can self-thermalize (and when not).}
A broad $3\!\leftrightarrow\!2$ kernel with sizable momentum transfer \emph{can} contribute to kineticization; in some UV completions the same diagrams that deplete number also exchange momentum efficiently across $p$.
However, near kinematic thresholds, for narrow resonances, or if the $3\!\to\!2$ matrix element is dominated by configurations with small momentum transfer, the induced momentum exchange per event is insufficient to maintain a Maxwellian shape on a Hubble time without the help of $2\!\to\!2$ elastic. These corners are precisely where a momentum-resolved check is warranted rather than assuming a single $T_d$.\\

\noindent \paragraph{Representative realizations.}
\begin{itemize}
\item \emph{Scalar cannibals with heavy mediator:} $\lambda \chi^6/\Lambda^2$ generates $3\!\to\!2$ at tree level (after EWSB-like symmetry breaking in the DS), while $\chi\chi\!\to\!\chi\chi$ proceeds via a heavy state of mass $M_d\!\sim\!\Lambda$; transport is suppressed by $M_d^{-4}$ and forward peaking in the nonrelativistic limit \cite{PappadopuloRuderman2016,FarinaPappadopuloRudermanTrevisan2016}.
\item \emph{Pseudo-Goldstone SIMP-like sectors:} derivative couplings enhance $3\!\to\!2$ at low $v$ but elastic is dominated by $t$-channel exchange with small scattering angles, suppressing $\sigma_{\rm mt}$ relative to $\sigma_{\rm el}$ \cite{Alanne:2020jwx}.
\end{itemize}

\subsection{Remarks on sufficient conditions}
Chemical \emph{and} kinetic equilibrium within the DS is ensured if at least one of the following holds:
\begin{enumerate}
\item Some DS elastic channel with light target (e.g.\ dark photon, light scalar) yields $\gamma_p^{\rm(DS)}\!\gg\!H$ throughout the chemically active epoch for all heavy species.
\item Number–changing processes themselves (e.g.\ broad-$s$ $2\!\leftrightarrow\!2$ or sufficiently frequent $3\!\leftrightarrow\!2$ with wide kinematics) also furnish efficient momentum exchange across populated $p$ without strong forward peaking.
\end{enumerate}
Absent these, DS chemistry can be maintained (shared $\mu_i$ at $T_d$) while DS kinetics fails (non-thermal $f_i$), even when the DS is fully secluded from the SM.

Note that dark freeze-out in a self-thermalized sector (with the \emph{assumption} of fast DS elastic) is developed in \cite{PappadopuloRuderman2016,FarinaPappadopuloRudermanTrevisan2016}. Momentum-exchange diagnostics via transport cross sections and FP/Langevin operators—directly applicable to DS kinetics—are discussed in \cite{BringmannHofmann2007,Ali-Haimoud:2018iiy}. Semi-annihilation frameworks providing chemistry with potentially weak DS elastic appear in $Z_3$-like models (e.g.\ \cite{DEramo:2010keq}); DS conversion/coannihilation analogues of \S\ref{subsec:DS-conversion} are analyzed in conversion-driven freeze-out contexts \cite{GarnyHeisig2017,GarnyHeisigHufnagelLuelfVogl2019}.


\section{Discussion and Conclusions}
\label{sec:discussion}

This paper has separated, both conceptually and operationally, two notions that are often conflated in early‐Universe dynamics: chemical equilibrium, which governs number densities through the slowest number–changing process, and kinetic equilibrium, which governs the shape of the momentum distribution through transport. The clean way to keep them apart is to track the rate that controls chemistry, $\Gamma_{\rm chem}$, alongside the transport–weighted momentum–exchange rate, $\gamma_p$, and to compare each to the Hubble rate over the momenta that actually carry number and energy. Because inelastic and elastic operators scale differently with couplings, propagators, angular structure, and thresholds, there is no generic implication in either direction. It is common to encounter parameter regions where $\Gamma_{\rm chem}\!\gg\!H$ while $\gamma_p\!\lesssim\!H$, or the converse, and it is equally common that the two ratios cross unity at different times so that chemical and kinetic decoupling proceed in different orders.

The Standard Model neutrino sector provides a precise illustration of these ideas. Around the MeV epoch, pair processes that control the neutrino chemical potentials thin first, while elastic scattering on $e^\pm$ continues to exchange momentum and energy into the era of $e^\pm$ annihilation. Momentum–resolved transport then predicts small, irreducible spectral distortions—largest at high comoving momenta—that cannot be absorbed into any single choice of $(T_\nu,\mu_\nu)$. The effect is minute but robust: it underlies the SM prediction $N_{\rm eff}\simeq 3.044$–$3.045$ and is a reminder that a species may have been fully equilibrated in the past and yet not retain an exactly thermal relic spectrum.

Heavy right‐handed neutrinos in strong‐washout leptogenesis illustrate the inverse separation. Inverse decays proportional to $|y_N|^2$ pin the abundance close to equilibrium at $T\!\sim\!M_N$, but the elastic processes that would efficiently thermalize momenta are $|y_N|^4$ and typically forward–peaked, so their transport rate is parametrically small. As a result, the momentum distribution can remain only partially relaxed even while the number density is chemically equilibrated. Any residual nonthermal structure simply redshifts after inelastic processes thin, and although strong washout remains intact, the detailed timing of washout and the efficiency factor can shift at the level relevant for precision studies.

Dark matter scenarios make the nontrivial ordering of decouplings especially transparent. When annihilation is concentrated by kinematics or resonances—near a narrow $s$–channel pole or into forbidden final states—chemical processing can remain efficient in a narrow band of velocities or in the high–velocity tail, even as elastic scattering on relativistic Standard Model targets is transport–suppressed by heavy propagators or forward peaking. In such regions one literally sees the kinetic decoupling occur first: elastic refilling fails, momentum–space “notches’’ open near the favored annihilation velocities, or the high–$p$ tail erodes, and only later does chemical freeze–out complete once the sculpted distribution can no longer sustain $\Gamma_{\rm chem}\!\gtrsim\!H$. The relic abundance inferred assuming kinetic equilibrium can then differ by factors of order unity, and in extreme cases by an order of magnitude, from the result obtained using the actual phase–space distribution.

Closely related hierarchies appear in coannihilation and conversion–driven settings, and in coscattering corridors. Conversions enforce chemical relations between a dark matter state and a slightly heavier partner, tying chemical potentials together and maintaining the equilibrium number ratio while the final depletion proceeds through the partner’s annihilations. If the direct elastic portal of the dark matter to the Standard Model is weak, the species is chemically tied to the bath through its partner but kinetically decoupled from it. In coscattering, an endothermic upscattering $\chi X\!\to\!\psi X'$ controls the abundance after annihilations thin; low–momentum modes fall below threshold and decouple first, and the depletion proceeds by nibbling away at the high–momentum tail. Here, too, kinetic decoupling precedes chemical decoupling in a way that is visible directly in the evolving $f(p)$.

All of these phenomena have analogues within secluded dark sectors at a temperature $T_d\neq T$. Chemical equilibration among dark species can be maintained by conversions, semi–annihilation, or $3\!\to\!2$ reactions, while intra–sector elastic scattering is too weak to enforce Maxwellian shapes. In such cases the sector has a well–defined chemical network at $T_d$ but not a common kinetic temperature, and the heavy component retains nonthermal features—self–heated shoulders from $3\!\to\!2$ or injection shoulders from semi–annihilation—even as the chemical constraints are satisfied. The textbook cannibal result $T_d\propto 1/\ln a$ relies on fast elastic self–scattering; when transport is marginal, the heavy component cools more slowly than the bath and a single–temperature closure is no longer reliable.

These examples motivate a simple methodological lesson. The reliable analysis is to compute $\Gamma_{\rm chem}(T)$ for the process that actually sets the abundance and to compute $\gamma_p(T)$ to the relevant bath using transport cross sections or a properly matched Fokker–Planck/Langevin description that reproduces both momentum and heat exchange. One should plot $\Gamma_{\rm chem}/H$ and $\gamma_p/H$ on the same axis through the epoch of interest and, whenever possible, resolve the momentum dependence of $\gamma_p/H$, because the highest–momentum modes typically fail first. The solver should match the physics: full collision integrals in momentum space are warranted when chemistry is momentum–selective—resonances, thresholds, coscattering—or when precision neutrino transport is at stake; Fokker–Planck/Langevin captures small–angle elastic dynamics efficiently and provides clean transport diagnostics; moment closures that evolve number density and a kinetic temperature are acceptable only while the spectrum remains close to Maxwellian. It is good practice to expose spectral evidence with snapshots of $p^2 f$ or $f/f_{\rm MB}$ and to compare thermal averages taken over the evolving distribution to their kinetic–equilibrium surrogates, because any bias there feeds directly into the inferred relic abundance or radiation density. In two–bath cosmologies one should also report the evolution of $T_d/T$, and, when intra–sector elastic is marginal, species–specific kinetic temperatures relative to $T_d$.

The phenomenological consequences of getting this right are concrete. In resonant and forbidden windows, early kinetic decoupling changes the relic density by order–one factors relative to equilibrium estimates, shifting the couplings that reproduce today’s abundance. In coannihilation and conversion–driven settings, the differences are often at the few–to–ten percent level but can become larger near thresholds. In the radiation sector, momentum–resolved neutrino transport underlies the small but precise shift in $N_{\rm eff}$ and fixes the baseline against which late injections or low reheating scenarios must be tested. Nonthermal late–time velocity distributions alter indirect–detection rates for velocity–dependent operators, including Sommerfeld–enhanced channels, and change the mapping to CMB energy–injection constraints. If kinetic decoupling occurs while a species is semi–relativistic, broadened or anisotropic distributions modify free–streaming scales and small–scale structure relative to thermal templates; in dark sectors with $T_d\neq T$, the combination of seclusion, self–heating, and nonthermal shapes imprints characteristic cutoffs.

The central lesson is straightforward: chemical equilibrium describes numerical abundances, while kinetic equilibrium captures distributional forms, and in our expanding Universe, the processes governing each typically decouple at different epochs. By maintaining both diagnostics in parallel—and invoking phase-space evolution when momentum-dependent effects become significant-one establishes the most reliable pathway to robust, testable cosmological predictions.

\begin{acknowledgments}
This work was inspired by a (correct) remark by Joshua Ruderman on the occasion of his lectures at the pre-SUSY2025 summer school; I acknowledge feedback from Joshua Ruderman, Pierce Giffin, and Grant Roberts. SP is partly supported by the U.S.\ Department of Energy grant number de-sc0010107.
\end{acknowledgments}

\bibliography{bib}

\begin{thebibliography}{39}%
\makeatletter
\providecommand \@ifxundefined [1]{%
 \@ifx{#1\undefined}
}%
\providecommand \@ifnum [1]{%
 \ifnum #1\expandafter \@firstoftwo
 \else \expandafter \@secondoftwo
 \fi
}%
\providecommand \@ifx [1]{%
 \ifx #1\expandafter \@firstoftwo
 \else \expandafter \@secondoftwo
 \fi
}%
\providecommand \natexlab [1]{#1}%
\providecommand \enquote  [1]{``#1''}%
\providecommand \bibnamefont  [1]{#1}%
\providecommand \bibfnamefont [1]{#1}%
\providecommand \citenamefont [1]{#1}%
\providecommand \href@noop [0]{\@secondoftwo}%
\providecommand \href [0]{\begingroup \@sanitize@url \@href}%
\providecommand \@href[1]{\@@startlink{#1}\@@href}%
\providecommand \@@href[1]{\endgroup#1\@@endlink}%
\providecommand \@sanitize@url [0]{\catcode `\\12\catcode `\$12\catcode `\&12\catcode `\#12\catcode `\^12\catcode `\_12\catcode `\%12\relax}%
\providecommand \@@startlink[1]{}%
\providecommand \@@endlink[0]{}%
\providecommand \url  [0]{\begingroup\@sanitize@url \@url }%
\providecommand \@url [1]{\endgroup\@href {#1}{\urlprefix }}%
\providecommand \urlprefix  [0]{URL }%
\providecommand \Eprint [0]{\href }%
\providecommand \doibase [0]{https://doi.org/}%
\providecommand \selectlanguage [0]{\@gobble}%
\providecommand \bibinfo  [0]{\@secondoftwo}%
\providecommand \bibfield  [0]{\@secondoftwo}%
\providecommand \translation [1]{[#1]}%
\providecommand \BibitemOpen [0]{}%
\providecommand \bibitemStop [0]{}%
\providecommand \bibitemNoStop [0]{.\EOS\space}%
\providecommand \EOS [0]{\spacefactor3000\relax}%
\providecommand \BibitemShut  [1]{\csname bibitem#1\endcsname}%
\let\auto@bib@innerbib\@empty
\bibitem [{\citenamefont {Peebles}(1993)}]{Peebles1993}%
  \BibitemOpen
  \bibfield  {author} {\bibinfo {author} {\bibfnamefont {P.~J.~E.}\ \bibnamefont {Peebles}},\ }\href@noop {} {\emph {\bibinfo {title} {{Principles of Physical Cosmology}}}}\ (\bibinfo  {publisher} {Princeton University Press},\ \bibinfo {address} {Princeton, NJ},\ \bibinfo {year} {1993})\BibitemShut {NoStop}%
\bibitem [{\citenamefont {Weinberg}(2008)}]{Weinberg2008}%
  \BibitemOpen
  \bibfield  {author} {\bibinfo {author} {\bibfnamefont {S.}~\bibnamefont {Weinberg}},\ }\href@noop {} {\emph {\bibinfo {title} {{Cosmology}}}}\ (\bibinfo  {publisher} {Oxford University Press},\ \bibinfo {address} {Oxford},\ \bibinfo {year} {2008})\BibitemShut {NoStop}%
\bibitem [{\citenamefont {Kolb}\ and\ \citenamefont {Turner}(1990)}]{Kolb1990}%
  \BibitemOpen
  \bibfield  {author} {\bibinfo {author} {\bibfnamefont {E.~W.}\ \bibnamefont {Kolb}}\ and\ \bibinfo {author} {\bibfnamefont {M.~S.}\ \bibnamefont {Turner}},\ }\href@noop {} {\emph {\bibinfo {title} {{The Early Universe}}}},\ \bibinfo {series} {Frontiers in Physics}, Vol.~\bibinfo {volume} {69}\ (\bibinfo  {publisher} {Addison-Wesley},\ \bibinfo {address} {Reading, MA},\ \bibinfo {year} {1990})\BibitemShut {NoStop}%
\bibitem [{\citenamefont {Shu}(1991)}]{Shu1991}%
  \BibitemOpen
  \bibfield  {author} {\bibinfo {author} {\bibfnamefont {F.~H.}\ \bibnamefont {Shu}},\ }\href@noop {} {\emph {\bibinfo {title} {{The Physics of Astrophysics. Volume 1: Radiation}}}}\ (\bibinfo  {publisher} {University Science Books},\ \bibinfo {address} {Mill Valley, CA},\ \bibinfo {year} {1991})\BibitemShut {NoStop}%
\bibitem [{\citenamefont {Bernstein}(1988)}]{Bernstein1988}%
  \BibitemOpen
  \bibfield  {author} {\bibinfo {author} {\bibfnamefont {J.}~\bibnamefont {Bernstein}},\ }\href@noop {} {\emph {\bibinfo {title} {{Kinetic Theory in the Expanding Universe}}}},\ Cambridge Monographs on Mathematical Physics\ (\bibinfo  {publisher} {Cambridge University Press},\ \bibinfo {address} {Cambridge},\ \bibinfo {year} {1988})\BibitemShut {NoStop}%
\bibitem [{\citenamefont {Dodelson}(2003)}]{Dodelson2003}%
  \BibitemOpen
  \bibfield  {author} {\bibinfo {author} {\bibfnamefont {S.}~\bibnamefont {Dodelson}},\ }\href@noop {} {\emph {\bibinfo {title} {{Modern Cosmology}}}}\ (\bibinfo  {publisher} {Academic Press},\ \bibinfo {address} {San Diego, CA},\ \bibinfo {year} {2003})\BibitemShut {NoStop}%
\bibitem [{\citenamefont {Pathria}\ and\ \citenamefont {Beale}(2011)}]{Pathria2011}%
  \BibitemOpen
  \bibfield  {author} {\bibinfo {author} {\bibfnamefont {R.~K.}\ \bibnamefont {Pathria}}\ and\ \bibinfo {author} {\bibfnamefont {P.~D.}\ \bibnamefont {Beale}},\ }\href@noop {} {\emph {\bibinfo {title} {{Statistical Mechanics}}}},\ \bibinfo {edition} {3rd}\ ed.\ (\bibinfo  {publisher} {Academic Press},\ \bibinfo {address} {Boston, MA},\ \bibinfo {year} {2011})\BibitemShut {NoStop}%
\bibitem [{\citenamefont {Gondolo}\ and\ \citenamefont {Gelmini}(1991)}]{Gondolo2012}%
  \BibitemOpen
  \bibfield  {author} {\bibinfo {author} {\bibfnamefont {P.}~\bibnamefont {Gondolo}}\ and\ \bibinfo {author} {\bibfnamefont {G.}~\bibnamefont {Gelmini}},\ }\bibfield  {title} {\bibinfo {title} {{Cosmic abundances of stable particles: Improved analysis}},\ }\href {https://doi.org/10.1016/0550-3213(91)90475-C} {\bibfield  {journal} {\bibinfo  {journal} {Nucl. Phys. B}\ }\textbf {\bibinfo {volume} {360}},\ \bibinfo {pages} {145} (\bibinfo {year} {1991})},\ \Eprint {https://arxiv.org/abs/hep-ph/9108207} {arXiv:hep-ph/9108207} \BibitemShut {NoStop}%
\bibitem [{\citenamefont {Dolgov}(2002)}]{Dolgov:2002wy}%
  \BibitemOpen
  \bibfield  {author} {\bibinfo {author} {\bibfnamefont {A.~D.}\ \bibnamefont {Dolgov}},\ }\bibfield  {title} {\bibinfo {title} {{Neutrinos in cosmology}},\ }\href {https://doi.org/10.1016/S0370-1573(02)00139-4} {\bibfield  {journal} {\bibinfo  {journal} {Phys. Rept.}\ }\textbf {\bibinfo {volume} {370}},\ \bibinfo {pages} {333} (\bibinfo {year} {2002})},\ \Eprint {https://arxiv.org/abs/hep-ph/0202122} {arXiv:hep-ph/0202122} \BibitemShut {NoStop}%
\bibitem [{\citenamefont {Bringmann}\ and\ \citenamefont {Hofmann}(2007)}]{BringmannHofmann2007}%
  \BibitemOpen
  \bibfield  {author} {\bibinfo {author} {\bibfnamefont {T.}~\bibnamefont {Bringmann}}\ and\ \bibinfo {author} {\bibfnamefont {S.}~\bibnamefont {Hofmann}},\ }\bibfield  {title} {\bibinfo {title} {Thermal decoupling of wimps from first principles},\ }\href {https://doi.org/10.1088/1475-7516/2007/04/016} {\bibfield  {journal} {\bibinfo  {journal} {JCAP}\ }\textbf {\bibinfo {volume} {2007}}\bibfield  {number} {\bibinfo  {number} { (04)},\ \bibinfo {pages} {016}},\ }\bibinfo {note} {erratum: JCAP 03 (2016) E02},\ \Eprint {https://arxiv.org/abs/hep-ph/0612238} {arXiv:hep-ph/0612238} \BibitemShut {NoStop}%
\bibitem [{\citenamefont {Boehm}\ \emph {et~al.}(2002)\citenamefont {Boehm}, \citenamefont {Riazuelo}, \citenamefont {Hansen},\ and\ \citenamefont {Schaeffer}}]{Boehm:2000gq}%
  \BibitemOpen
  \bibfield  {author} {\bibinfo {author} {\bibfnamefont {C.}~\bibnamefont {Boehm}}, \bibinfo {author} {\bibfnamefont {A.}~\bibnamefont {Riazuelo}}, \bibinfo {author} {\bibfnamefont {S.~H.}\ \bibnamefont {Hansen}},\ and\ \bibinfo {author} {\bibfnamefont {R.}~\bibnamefont {Schaeffer}},\ }\bibfield  {title} {\bibinfo {title} {Interacting dark matter disguised as warm dark matter},\ }\href {https://doi.org/10.1103/PhysRevD.66.083505} {\bibfield  {journal} {\bibinfo  {journal} {Phys. Rev. D}\ }\textbf {\bibinfo {volume} {66}},\ \bibinfo {pages} {083505} (\bibinfo {year} {2002})},\ \Eprint {https://arxiv.org/abs/astro-ph/0012194} {arXiv:astro-ph/0012194} \BibitemShut {NoStop}%
\bibitem [{\citenamefont {Du}\ \emph {et~al.}(2022)\citenamefont {Du}, \citenamefont {Huang}, \citenamefont {Li}, \citenamefont {Li},\ and\ \citenamefont {Yu}}]{Du:2021jcj}%
  \BibitemOpen
  \bibfield  {author} {\bibinfo {author} {\bibfnamefont {Y.}~\bibnamefont {Du}}, \bibinfo {author} {\bibfnamefont {F.}~\bibnamefont {Huang}}, \bibinfo {author} {\bibfnamefont {H.-L.}\ \bibnamefont {Li}}, \bibinfo {author} {\bibfnamefont {Y.-Z.}\ \bibnamefont {Li}},\ and\ \bibinfo {author} {\bibfnamefont {J.-H.}\ \bibnamefont {Yu}},\ }\bibfield  {title} {\bibinfo {title} {{Revisiting dark matter freeze-in and freeze-out through phase-space distribution}},\ }\href {https://doi.org/10.1088/1475-7516/2022/04/012} {\bibfield  {journal} {\bibinfo  {journal} {JCAP}\ }\textbf {\bibinfo {volume} {04}}\bibfield  {number} {\bibinfo  {number} { (04)},\ \bibinfo {pages} {012}},\ }\Eprint {https://arxiv.org/abs/2111.01267} {arXiv:2111.01267 [hep-ph]} \BibitemShut {NoStop}%
\bibitem [{\citenamefont {Hofmann}\ \emph {et~al.}(2001)\citenamefont {Hofmann}, \citenamefont {Schwarz},\ and\ \citenamefont {Stoecker}}]{Hofmann:2001bi}%
  \BibitemOpen
  \bibfield  {author} {\bibinfo {author} {\bibfnamefont {S.}~\bibnamefont {Hofmann}}, \bibinfo {author} {\bibfnamefont {D.~J.}\ \bibnamefont {Schwarz}},\ and\ \bibinfo {author} {\bibfnamefont {H.}~\bibnamefont {Stoecker}},\ }\bibfield  {title} {\bibinfo {title} {Damping scales of neutralino cold dark matter},\ }\href {https://doi.org/10.1103/PhysRevD.64.083507} {\bibfield  {journal} {\bibinfo  {journal} {Phys. Rev. D}\ }\textbf {\bibinfo {volume} {64}},\ \bibinfo {pages} {083507} (\bibinfo {year} {2001})},\ \Eprint {https://arxiv.org/abs/astro-ph/0104173} {arXiv:astro-ph/0104173} \BibitemShut {NoStop}%
\bibitem [{\citenamefont {Vogelsberger}\ \emph {et~al.}(2016)\citenamefont {Vogelsberger}, \citenamefont {Zavala}, \citenamefont {Cyr-Racine}, \citenamefont {Pfrommer}, \citenamefont {Bringmann},\ and\ \citenamefont {Sigurdson}}]{Vogelsberger:2015gpr}%
  \BibitemOpen
  \bibfield  {author} {\bibinfo {author} {\bibfnamefont {M.}~\bibnamefont {Vogelsberger}}, \bibinfo {author} {\bibfnamefont {J.}~\bibnamefont {Zavala}}, \bibinfo {author} {\bibfnamefont {F.-Y.}\ \bibnamefont {Cyr-Racine}}, \bibinfo {author} {\bibfnamefont {C.}~\bibnamefont {Pfrommer}}, \bibinfo {author} {\bibfnamefont {T.}~\bibnamefont {Bringmann}},\ and\ \bibinfo {author} {\bibfnamefont {K.}~\bibnamefont {Sigurdson}},\ }\bibfield  {title} {\bibinfo {title} {Ethos – an effective theory of structure formation: dark matter physics as a possible explanation of the small-scale cdm problems},\ }\href {https://doi.org/10.1093/mnras/stw713} {\bibfield  {journal} {\bibinfo  {journal} {Mon. Not. Roy. Astron. Soc.}\ }\textbf {\bibinfo {volume} {460}},\ \bibinfo {pages} {1399} (\bibinfo {year} {2016})},\ \Eprint {https://arxiv.org/abs/1512.05349} {arXiv:1512.05349 [astro-ph.CO]} \BibitemShut {NoStop}%
\bibitem [{\citenamefont {Slatyer}\ and\ \citenamefont {Wu}(2018)}]{Slatyer:2018aqg}%
  \BibitemOpen
  \bibfield  {author} {\bibinfo {author} {\bibfnamefont {T.~R.}\ \bibnamefont {Slatyer}}\ and\ \bibinfo {author} {\bibfnamefont {C.-L.}\ \bibnamefont {Wu}},\ }\bibfield  {title} {\bibinfo {title} {General constraints on dark matter decay from the cosmic microwave background},\ }\href {https://doi.org/10.1103/PhysRevD.98.023013} {\bibfield  {journal} {\bibinfo  {journal} {Phys. Rev. D}\ }\textbf {\bibinfo {volume} {98}},\ \bibinfo {pages} {023013} (\bibinfo {year} {2018})},\ \Eprint {https://arxiv.org/abs/1803.09734} {arXiv:1803.09734 [astro-ph.CO]} \BibitemShut {NoStop}%
\bibitem [{\citenamefont {Binder}\ \emph {et~al.}(2021)\citenamefont {Binder}, \citenamefont {Bringmann}, \citenamefont {Gustafsson},\ and\ \citenamefont {Hryczuk}}]{Binder:2021bmg}%
  \BibitemOpen
  \bibfield  {author} {\bibinfo {author} {\bibfnamefont {T.}~\bibnamefont {Binder}}, \bibinfo {author} {\bibfnamefont {T.}~\bibnamefont {Bringmann}}, \bibinfo {author} {\bibfnamefont {M.}~\bibnamefont {Gustafsson}},\ and\ \bibinfo {author} {\bibfnamefont {A.}~\bibnamefont {Hryczuk}},\ }\bibfield  {title} {\bibinfo {title} {{Dark matter relic abundance beyond kinetic equilibrium}},\ }\href {https://doi.org/10.1140/epjc/s10052-021-09357-5} {\bibfield  {journal} {\bibinfo  {journal} {Eur. Phys. J. C}\ }\textbf {\bibinfo {volume} {81}},\ \bibinfo {pages} {577} (\bibinfo {year} {2021})},\ \Eprint {https://arxiv.org/abs/2103.01944} {arXiv:2103.01944 [hep-ph]} \BibitemShut {NoStop}%
\bibitem [{\citenamefont {Bode}\ \emph {et~al.}(2001)\citenamefont {Bode}, \citenamefont {Ostriker},\ and\ \citenamefont {Turok}}]{Bode:2000gq}%
  \BibitemOpen
  \bibfield  {author} {\bibinfo {author} {\bibfnamefont {P.}~\bibnamefont {Bode}}, \bibinfo {author} {\bibfnamefont {J.~P.}\ \bibnamefont {Ostriker}},\ and\ \bibinfo {author} {\bibfnamefont {N.}~\bibnamefont {Turok}},\ }\bibfield  {title} {\bibinfo {title} {Halo formation in warm dark matter models},\ }\href {https://doi.org/10.1086/321541} {\bibfield  {journal} {\bibinfo  {journal} {Astrophys. J.}\ }\textbf {\bibinfo {volume} {556}},\ \bibinfo {pages} {93} (\bibinfo {year} {2001})},\ \Eprint {https://arxiv.org/abs/astro-ph/0010389} {arXiv:astro-ph/0010389} \BibitemShut {NoStop}%
\bibitem [{\citenamefont {Chu}\ \emph {et~al.}(2012)\citenamefont {Chu}, \citenamefont {Hambye},\ and\ \citenamefont {Tytgat}}]{Chu:2011be}%
  \BibitemOpen
  \bibfield  {author} {\bibinfo {author} {\bibfnamefont {X.}~\bibnamefont {Chu}}, \bibinfo {author} {\bibfnamefont {T.}~\bibnamefont {Hambye}},\ and\ \bibinfo {author} {\bibfnamefont {M.~H.~G.}\ \bibnamefont {Tytgat}},\ }\bibfield  {title} {\bibinfo {title} {Thermal relic dark matter in scenarios with large annihilation cross sections},\ }\href {https://doi.org/10.1088/1475-7516/2012/05/034} {\bibfield  {journal} {\bibinfo  {journal} {JCAP}\ }\textbf {\bibinfo {volume} {05}},\ \bibinfo {pages} {034}},\ \Eprint {https://arxiv.org/abs/1112.0493} {arXiv:1112.0493 [hep-ph]} \BibitemShut {NoStop}%
\bibitem [{\citenamefont {Feng}\ \emph {et~al.}(2009)\citenamefont {Feng}, \citenamefont {Kaplinghat}, \citenamefont {Tu},\ and\ \citenamefont {Yu}}]{Feng2008}%
  \BibitemOpen
  \bibfield  {author} {\bibinfo {author} {\bibfnamefont {J.~L.}\ \bibnamefont {Feng}}, \bibinfo {author} {\bibfnamefont {M.}~\bibnamefont {Kaplinghat}}, \bibinfo {author} {\bibfnamefont {H.}~\bibnamefont {Tu}},\ and\ \bibinfo {author} {\bibfnamefont {H.-B.}\ \bibnamefont {Yu}},\ }\bibfield  {title} {\bibinfo {title} {{Hidden charged dark matter}},\ }\href {https://doi.org/10.1088/1475-7516/2009/07/004} {\bibfield  {journal} {\bibinfo  {journal} {JCAP}\ }\textbf {\bibinfo {volume} {07}},\ \bibinfo {pages} {004}},\ \Eprint {https://arxiv.org/abs/0905.3039} {arXiv:0905.3039 [hep-ph]} \BibitemShut {NoStop}%
\bibitem [{\citenamefont {Hryczuk}\ and\ \citenamefont {Laletin}(2022)}]{Hryczuk:2022gay}%
  \BibitemOpen
  \bibfield  {author} {\bibinfo {author} {\bibfnamefont {A.}~\bibnamefont {Hryczuk}}\ and\ \bibinfo {author} {\bibfnamefont {M.}~\bibnamefont {Laletin}},\ }\bibfield  {title} {\bibinfo {title} {{Impact of dark matter self-scattering on its relic abundance}},\ }\href {https://doi.org/10.1103/PhysRevD.106.023007} {\bibfield  {journal} {\bibinfo  {journal} {Phys. Rev. D}\ }\textbf {\bibinfo {volume} {106}},\ \bibinfo {pages} {023007} (\bibinfo {year} {2022})},\ \Eprint {https://arxiv.org/abs/2204.07078} {arXiv:2204.07078 [hep-ph]} \BibitemShut {NoStop}%
\bibitem [{\citenamefont {Binder}\ \emph {et~al.}(2017)\citenamefont {Binder}, \citenamefont {Bringmann}, \citenamefont {Gustafsson},\ and\ \citenamefont {Hryczuk}}]{Binder:2017rgn}%
  \BibitemOpen
  \bibfield  {author} {\bibinfo {author} {\bibfnamefont {T.}~\bibnamefont {Binder}}, \bibinfo {author} {\bibfnamefont {T.}~\bibnamefont {Bringmann}}, \bibinfo {author} {\bibfnamefont {M.}~\bibnamefont {Gustafsson}},\ and\ \bibinfo {author} {\bibfnamefont {A.}~\bibnamefont {Hryczuk}},\ }\bibfield  {title} {\bibinfo {title} {{Early kinetic decoupling of dark matter: When the standard way of calculating the thermal relic density fails}},\ }\href {https://doi.org/10.1103/PhysRevD.96.115010} {\bibfield  {journal} {\bibinfo  {journal} {Phys. Rev. D}\ }\textbf {\bibinfo {volume} {96}},\ \bibinfo {pages} {115010} (\bibinfo {year} {2017})},\ \Eprint {https://arxiv.org/abs/1706.07433} {arXiv:1706.07433 [hep-ph]} \BibitemShut {NoStop}%
\bibitem [{\citenamefont {D'Eramo}\ and\ \citenamefont {Thaler}(2010)}]{DEramo:2010keq}%
  \BibitemOpen
  \bibfield  {author} {\bibinfo {author} {\bibfnamefont {F.}~\bibnamefont {D'Eramo}}\ and\ \bibinfo {author} {\bibfnamefont {J.}~\bibnamefont {Thaler}},\ }\bibfield  {title} {\bibinfo {title} {{Semi-annihilation of Dark Matter}},\ }\href {https://doi.org/10.1007/JHEP06(2010)109} {\bibfield  {journal} {\bibinfo  {journal} {JHEP}\ }\textbf {\bibinfo {volume} {06}},\ \bibinfo {pages} {109}},\ \Eprint {https://arxiv.org/abs/1003.5912} {arXiv:1003.5912 [hep-ph]} \BibitemShut {NoStop}%
\bibitem [{\citenamefont {Pappadopulo}\ \emph {et~al.}(2016)\citenamefont {Pappadopulo}, \citenamefont {Ruderman},\ and\ \citenamefont {Trevisan}}]{PappadopuloRuderman2016}%
  \BibitemOpen
  \bibfield  {author} {\bibinfo {author} {\bibfnamefont {D.}~\bibnamefont {Pappadopulo}}, \bibinfo {author} {\bibfnamefont {J.~T.}\ \bibnamefont {Ruderman}},\ and\ \bibinfo {author} {\bibfnamefont {G.}~\bibnamefont {Trevisan}},\ }\bibfield  {title} {\bibinfo {title} {Dark matter freeze-out in a nonrelativistic sector},\ }\href {https://doi.org/10.1103/PhysRevD.94.035005} {\bibfield  {journal} {\bibinfo  {journal} {Phys. Rev. D}\ }\textbf {\bibinfo {volume} {94}},\ \bibinfo {pages} {035005} (\bibinfo {year} {2016})},\ \Eprint {https://arxiv.org/abs/1602.04219} {1602.04219} \BibitemShut {NoStop}%
\bibitem [{\citenamefont {Chatterjee}\ and\ \citenamefont {Hryczuk}(2025)}]{Chatterjee:2025vdz}%
  \BibitemOpen
  \bibfield  {author} {\bibinfo {author} {\bibfnamefont {S.}~\bibnamefont {Chatterjee}}\ and\ \bibinfo {author} {\bibfnamefont {A.}~\bibnamefont {Hryczuk}},\ }\bibfield  {title} {\bibinfo {title} {{Conversions in two-component dark sectors: a phase space level analysis}},\ }\href {https://doi.org/10.1007/JHEP07(2025)279} {\bibfield  {journal} {\bibinfo  {journal} {JHEP}\ }\textbf {\bibinfo {volume} {07}},\ \bibinfo {pages} {279}},\ \Eprint {https://arxiv.org/abs/2502.08725} {arXiv:2502.08725 [hep-ph]} \BibitemShut {NoStop}%
\bibitem [{\citenamefont {Mangano}\ \emph {et~al.}(2005)\citenamefont {Mangano}, \citenamefont {Miele}, \citenamefont {Pastor}, \citenamefont {Pinto}, \citenamefont {Pisanti},\ and\ \citenamefont {Serpico}}]{Mangano2005}%
  \BibitemOpen
  \bibfield  {author} {\bibinfo {author} {\bibfnamefont {G.}~\bibnamefont {Mangano}}, \bibinfo {author} {\bibfnamefont {G.}~\bibnamefont {Miele}}, \bibinfo {author} {\bibfnamefont {S.}~\bibnamefont {Pastor}}, \bibinfo {author} {\bibfnamefont {T.}~\bibnamefont {Pinto}}, \bibinfo {author} {\bibfnamefont {O.}~\bibnamefont {Pisanti}},\ and\ \bibinfo {author} {\bibfnamefont {P.~D.}\ \bibnamefont {Serpico}},\ }\bibfield  {title} {\bibinfo {title} {Relic neutrino decoupling including flavor oscillations},\ }\href {https://doi.org/10.1016/j.nuclphysb.2005.09.041} {\bibfield  {journal} {\bibinfo  {journal} {Nucl. Phys. B}\ }\textbf {\bibinfo {volume} {729}},\ \bibinfo {pages} {221} (\bibinfo {year} {2005})},\ \Eprint {https://arxiv.org/abs/hep-ph/0506164} {hep-ph/0506164} \BibitemShut {NoStop}%
\bibitem [{\citenamefont {Escudero}(2019)}]{Escudero2019}%
  \BibitemOpen
  \bibfield  {author} {\bibinfo {author} {\bibfnamefont {M.}~\bibnamefont {Escudero}},\ }\bibfield  {title} {\bibinfo {title} {Neutrino decoupling beyond the standard model: Cmb constraints on the dark matter mass with a fast and precise $n_{\rm eff}$ evaluation},\ }\href {https://doi.org/10.1088/1475-7516/2019/02/007} {\bibfield  {journal} {\bibinfo  {journal} {JCAP}\ }\textbf {\bibinfo {volume} {2019}}\bibfield  {number} {\bibinfo  {number} { (02)},\ \bibinfo {pages} {007}},\ }\Eprint {https://arxiv.org/abs/1812.05605} {arXiv:1812.05605 [hep-ph]} \BibitemShut {NoStop}%
\bibitem [{\citenamefont {Akita}\ and\ \citenamefont {Yamaguchi}(2020)}]{AkitaYamaguchi2020}%
  \BibitemOpen
  \bibfield  {author} {\bibinfo {author} {\bibfnamefont {K.}~\bibnamefont {Akita}}\ and\ \bibinfo {author} {\bibfnamefont {M.}~\bibnamefont {Yamaguchi}},\ }\bibfield  {title} {\bibinfo {title} {A precision calculation of relic neutrino decoupling},\ }\href {https://doi.org/10.1088/1475-7516/2020/08/012} {\bibfield  {journal} {\bibinfo  {journal} {JCAP}\ }\textbf {\bibinfo {volume} {08}},\ \bibinfo {pages} {012}},\ \Eprint {https://arxiv.org/abs/2005.07047} {2005.07047} \BibitemShut {NoStop}%
\bibitem [{\citenamefont {Cielo}\ \emph {et~al.}(2023)\citenamefont {Cielo}, \citenamefont {Escudero}, \citenamefont {Mangano},\ and\ \citenamefont {Pisanti}}]{CieloEtAl2023}%
  \BibitemOpen
  \bibfield  {author} {\bibinfo {author} {\bibfnamefont {M.}~\bibnamefont {Cielo}}, \bibinfo {author} {\bibfnamefont {M.}~\bibnamefont {Escudero}}, \bibinfo {author} {\bibfnamefont {G.}~\bibnamefont {Mangano}},\ and\ \bibinfo {author} {\bibfnamefont {O.}~\bibnamefont {Pisanti}},\ }\bibfield  {title} {\bibinfo {title} {{Neff in the Standard Model at NLO is 3.043}},\ }\href {https://doi.org/10.1103/PhysRevD.108.L121301} {\bibfield  {journal} {\bibinfo  {journal} {Phys. Rev. D}\ }\textbf {\bibinfo {volume} {108}},\ \bibinfo {pages} {L121301} (\bibinfo {year} {2023})},\ \Eprint {https://arxiv.org/abs/2306.05460} {arXiv:2306.05460 [hep-ph]} \BibitemShut {NoStop}%
\bibitem [{\citenamefont {Giudice}\ \emph {et~al.}(2004)\citenamefont {Giudice}, \citenamefont {Notari}, \citenamefont {Raidal}, \citenamefont {Riotto},\ and\ \citenamefont {Strumia}}]{Giudice:2003jh}%
  \BibitemOpen
  \bibfield  {author} {\bibinfo {author} {\bibfnamefont {G.~F.}\ \bibnamefont {Giudice}}, \bibinfo {author} {\bibfnamefont {A.}~\bibnamefont {Notari}}, \bibinfo {author} {\bibfnamefont {M.}~\bibnamefont {Raidal}}, \bibinfo {author} {\bibfnamefont {A.}~\bibnamefont {Riotto}},\ and\ \bibinfo {author} {\bibfnamefont {A.}~\bibnamefont {Strumia}},\ }\bibfield  {title} {\bibinfo {title} {Towards a complete theory of thermal leptogenesis in the {SM} and {MSSM}},\ }\href {https://doi.org/10.1016/j.nuclphysb.2004.02.019} {\bibfield  {journal} {\bibinfo  {journal} {Nucl. Phys. B}\ }\textbf {\bibinfo {volume} {685}},\ \bibinfo {pages} {89} (\bibinfo {year} {2004})},\ \Eprint {https://arxiv.org/abs/hep-ph/0310123} {arXiv:hep-ph/0310123} \BibitemShut {NoStop}%
\bibitem [{\citenamefont {Beneke}\ \emph {et~al.}(2010)\citenamefont {Beneke}, \citenamefont {Garbrecht}, \citenamefont {Herranen},\ and\ \citenamefont {Schwaller}}]{Beneke:2010wd}%
  \BibitemOpen
  \bibfield  {author} {\bibinfo {author} {\bibfnamefont {M.}~\bibnamefont {Beneke}}, \bibinfo {author} {\bibfnamefont {B.}~\bibnamefont {Garbrecht}}, \bibinfo {author} {\bibfnamefont {M.}~\bibnamefont {Herranen}},\ and\ \bibinfo {author} {\bibfnamefont {P.}~\bibnamefont {Schwaller}},\ }\bibfield  {title} {\bibinfo {title} {Finite number density corrections to leptogenesis},\ }\href {https://doi.org/10.1016/j.nuclphysb.2010.05.003} {\bibfield  {journal} {\bibinfo  {journal} {Nucl. Phys. B}\ }\textbf {\bibinfo {volume} {838}},\ \bibinfo {pages} {1} (\bibinfo {year} {2010})},\ \Eprint {https://arxiv.org/abs/1002.1326} {arXiv:1002.1326 [hep-ph]} \BibitemShut {NoStop}%
\bibitem [{\citenamefont {Ali-Ha{\"\i}moud}(2019)}]{Ali-Haimoud:2018iiy}%
  \BibitemOpen
  \bibfield  {author} {\bibinfo {author} {\bibfnamefont {Y.}~\bibnamefont {Ali-Ha{\"\i}moud}},\ }\bibfield  {title} {\bibinfo {title} {{Boltzmann-Fokker-Planck formalism for dark-matter--baryon scattering}},\ }\href {https://doi.org/10.1103/PhysRevD.99.023523} {\bibfield  {journal} {\bibinfo  {journal} {Phys. Rev. D}\ }\textbf {\bibinfo {volume} {99}},\ \bibinfo {pages} {023523} (\bibinfo {year} {2019})},\ \Eprint {https://arxiv.org/abs/1811.09903} {arXiv:1811.09903 [astro-ph.CO]} \BibitemShut {NoStop}%
\bibitem [{\citenamefont {Griest}\ and\ \citenamefont {Seckel}(1991)}]{GriestSeckel1991}%
  \BibitemOpen
  \bibfield  {author} {\bibinfo {author} {\bibfnamefont {K.}~\bibnamefont {Griest}}\ and\ \bibinfo {author} {\bibfnamefont {D.}~\bibnamefont {Seckel}},\ }\bibfield  {title} {\bibinfo {title} {{Three exceptions in the calculation of relic abundances}},\ }\href {https://doi.org/10.1103/PhysRevD.43.3191} {\bibfield  {journal} {\bibinfo  {journal} {Phys. Rev. D}\ }\textbf {\bibinfo {volume} {43}},\ \bibinfo {pages} {3191} (\bibinfo {year} {1991})}\BibitemShut {NoStop}%
\bibitem [{\citenamefont {Garny}\ \emph {et~al.}(2017)\citenamefont {Garny}, \citenamefont {Heisig}, \citenamefont {L{\"u}lf},\ and\ \citenamefont {Vogl}}]{GarnyHeisig2017}%
  \BibitemOpen
  \bibfield  {author} {\bibinfo {author} {\bibfnamefont {M.}~\bibnamefont {Garny}}, \bibinfo {author} {\bibfnamefont {J.}~\bibnamefont {Heisig}}, \bibinfo {author} {\bibfnamefont {B.}~\bibnamefont {L{\"u}lf}},\ and\ \bibinfo {author} {\bibfnamefont {S.}~\bibnamefont {Vogl}},\ }\bibfield  {title} {\bibinfo {title} {Coannihilation without chemical equilibrium},\ }\href {https://doi.org/10.1103/PhysRevD.96.103521} {\bibfield  {journal} {\bibinfo  {journal} {Phys. Rev. D}\ }\textbf {\bibinfo {volume} {96}},\ \bibinfo {pages} {103521} (\bibinfo {year} {2017})},\ \Eprint {https://arxiv.org/abs/1705.09292} {arXiv:1705.09292 [hep-ph]} \BibitemShut {NoStop}%
\bibitem [{\citenamefont {Garny}\ \emph {et~al.}(2019)\citenamefont {Garny}, \citenamefont {Heisig}, \citenamefont {Hufnagel}, \citenamefont {L{\"u}lf},\ and\ \citenamefont {Vogl}}]{GarnyHeisigHufnagelLuelfVogl2019}%
  \BibitemOpen
  \bibfield  {author} {\bibinfo {author} {\bibfnamefont {M.}~\bibnamefont {Garny}}, \bibinfo {author} {\bibfnamefont {J.}~\bibnamefont {Heisig}}, \bibinfo {author} {\bibfnamefont {M.}~\bibnamefont {Hufnagel}}, \bibinfo {author} {\bibfnamefont {B.}~\bibnamefont {L{\"u}lf}},\ and\ \bibinfo {author} {\bibfnamefont {S.}~\bibnamefont {Vogl}},\ }\bibfield  {title} {\bibinfo {title} {{Conversion-driven freeze-out: Dark matter genesis beyond the WIMP paradigm}},\ }\href {https://doi.org/10.22323/1.347.0092} {\bibfield  {journal} {\bibinfo  {journal} {PoS}\ }\textbf {\bibinfo {volume} {CORFU2018}},\ \bibinfo {pages} {092} (\bibinfo {year} {2019})},\ \Eprint {https://arxiv.org/abs/1904.00238} {arXiv:1904.00238 [hep-ph]} \BibitemShut {NoStop}%
\bibitem [{\citenamefont {D'Agnolo}\ \emph {et~al.}(2017)\citenamefont {D'Agnolo}, \citenamefont {Pappadopulo},\ and\ \citenamefont {Ruderman}}]{DAgnoloPappadopuloRuderman2017}%
  \BibitemOpen
  \bibfield  {author} {\bibinfo {author} {\bibfnamefont {R.~T.}\ \bibnamefont {D'Agnolo}}, \bibinfo {author} {\bibfnamefont {D.}~\bibnamefont {Pappadopulo}},\ and\ \bibinfo {author} {\bibfnamefont {J.~T.}\ \bibnamefont {Ruderman}},\ }\bibfield  {title} {\bibinfo {title} {Fourth exception in the calculation of relic abundances},\ }\href {https://doi.org/10.1103/PhysRevLett.119.061102} {\bibfield  {journal} {\bibinfo  {journal} {Phys. Rev. Lett.}\ }\textbf {\bibinfo {volume} {119}},\ \bibinfo {pages} {061102} (\bibinfo {year} {2017})},\ \Eprint {https://arxiv.org/abs/1705.08450} {arXiv:1705.08450 [hep-ph]} \BibitemShut {NoStop}%
\bibitem [{\citenamefont {Ala-Mattinen}\ \emph {et~al.}(2022)\citenamefont {Ala-Mattinen}, \citenamefont {Heikinheimo}, \citenamefont {Kainulainen},\ and\ \citenamefont {Tuominen}}]{Ala-Mattinen:2022nnh}%
  \BibitemOpen
  \bibfield  {author} {\bibinfo {author} {\bibfnamefont {K.}~\bibnamefont {Ala-Mattinen}}, \bibinfo {author} {\bibfnamefont {M.}~\bibnamefont {Heikinheimo}}, \bibinfo {author} {\bibfnamefont {K.}~\bibnamefont {Kainulainen}},\ and\ \bibinfo {author} {\bibfnamefont {K.}~\bibnamefont {Tuominen}},\ }\bibfield  {title} {\bibinfo {title} {{Momentum distributions of cosmic relics: Improved analysis}},\ }\href {https://doi.org/10.1103/PhysRevD.105.123005} {\bibfield  {journal} {\bibinfo  {journal} {Phys. Rev. D}\ }\textbf {\bibinfo {volume} {105}},\ \bibinfo {pages} {123005} (\bibinfo {year} {2022})},\ \Eprint {https://arxiv.org/abs/2201.06456} {arXiv:2201.06456 [hep-ph]} \BibitemShut {NoStop}%
\bibitem [{\citenamefont {Aboubrahim}\ \emph {et~al.}(2023)\citenamefont {Aboubrahim}, \citenamefont {Klasen},\ and\ \citenamefont {Wiggering}}]{Aboubrahim:2023pyr}%
  \BibitemOpen
  \bibfield  {author} {\bibinfo {author} {\bibfnamefont {A.}~\bibnamefont {Aboubrahim}}, \bibinfo {author} {\bibfnamefont {M.}~\bibnamefont {Klasen}},\ and\ \bibinfo {author} {\bibfnamefont {L.~P.}\ \bibnamefont {Wiggering}},\ }\bibfield  {title} {\bibinfo {title} {{Forbidden dark matter annihilation into leptons with full collision terms}},\ }\href {https://doi.org/10.1088/1475-7516/2023/08/075} {\bibfield  {journal} {\bibinfo  {journal} {JCAP}\ }\textbf {\bibinfo {volume} {08}},\ \bibinfo {pages} {075}},\ \Eprint {https://arxiv.org/abs/2306.07753} {arXiv:2306.07753 [hep-ph]} \BibitemShut {NoStop}%
\bibitem [{\citenamefont {Farina}\ \emph {et~al.}(2016)\citenamefont {Farina}, \citenamefont {Pappadopulo}, \citenamefont {Ruderman},\ and\ \citenamefont {Trevisan}}]{FarinaPappadopuloRudermanTrevisan2016}%
  \BibitemOpen
  \bibfield  {author} {\bibinfo {author} {\bibfnamefont {M.}~\bibnamefont {Farina}}, \bibinfo {author} {\bibfnamefont {D.}~\bibnamefont {Pappadopulo}}, \bibinfo {author} {\bibfnamefont {J.~T.}\ \bibnamefont {Ruderman}},\ and\ \bibinfo {author} {\bibfnamefont {G.}~\bibnamefont {Trevisan}},\ }\bibfield  {title} {\bibinfo {title} {Phases of cannibal dark matter},\ }\href {https://doi.org/10.1007/JHEP12(2016)039} {\bibfield  {journal} {\bibinfo  {journal} {JHEP}\ }\textbf {\bibinfo {volume} {12}},\ \bibinfo {pages} {039}},\ \Eprint {https://arxiv.org/abs/1607.03108} {1607.03108} \BibitemShut {NoStop}%
\bibitem [{\citenamefont {Alanne}\ \emph {et~al.}(2020)\citenamefont {Alanne}, \citenamefont {Benincasa}, \citenamefont {Heikinheimo}, \citenamefont {Kannike}, \citenamefont {Keus}, \citenamefont {Koivunen},\ and\ \citenamefont {Tuominen}}]{Alanne:2020jwx}%
  \BibitemOpen
  \bibfield  {author} {\bibinfo {author} {\bibfnamefont {T.}~\bibnamefont {Alanne}}, \bibinfo {author} {\bibfnamefont {N.}~\bibnamefont {Benincasa}}, \bibinfo {author} {\bibfnamefont {M.}~\bibnamefont {Heikinheimo}}, \bibinfo {author} {\bibfnamefont {K.}~\bibnamefont {Kannike}}, \bibinfo {author} {\bibfnamefont {V.}~\bibnamefont {Keus}}, \bibinfo {author} {\bibfnamefont {N.}~\bibnamefont {Koivunen}},\ and\ \bibinfo {author} {\bibfnamefont {K.}~\bibnamefont {Tuominen}},\ }\bibfield  {title} {\bibinfo {title} {{Pseudo-Goldstone dark matter: gravitational waves and direct-detection blind spots}},\ }\href {https://doi.org/10.1007/JHEP10(2020)080} {\bibfield  {journal} {\bibinfo  {journal} {JHEP}\ }\textbf {\bibinfo {volume} {10}},\ \bibinfo {pages} {080}},\ \Eprint {https://arxiv.org/abs/2008.09605} {arXiv:2008.09605 [hep-ph]} \BibitemShut {NoStop}%
\end{thebibliography}%

\end{document}